\documentclass[a4paper,preprint,showpacs,superscriptaddress]{revtex4}
\usepackage[version=3]{mhchem}
\usepackage{float,fancyhdr,amsmath,graphicx,subfig,epsfig,color,setspace,url}
\usepackage{appendix}
\begin{document}

\title{Dissociative Electron Attachment to Polyatomic Molecules - IV : Methane }
\author{N. Bhargava Ram}
\email[]{nbhargavaram@tifr.res.in}
\affiliation{Tata Institute of Fundamental Research, Mumbai 400005, India}

\author{E. Krishnakumar}
\email[]{ekkumar@tifr.res.in}
\affiliation{Tata Institute of Fundamental Research,  Mumbai 400005, India}

\begin{abstract}

In this paper, we discuss the dissociative electron attachment process in Methane. Kinetic energy and angular distributions of \ce{H-} and \ce{CH2^{-}} fragment anions across the broad resonance about 10 eV is reported. Angular distribution of \ce{H-} ions changes from being perpendicular to forward-backward scattering across the resonance. Possibility of Jahn-Teller effect on excitation of the triply degenerate \ce{1t2} molecular orbital is inferred.   
\end{abstract}


\maketitle

\section{Introduction}

\begin{figure}[!htbp]
\centering
\includegraphics[width=0.7\columnwidth]{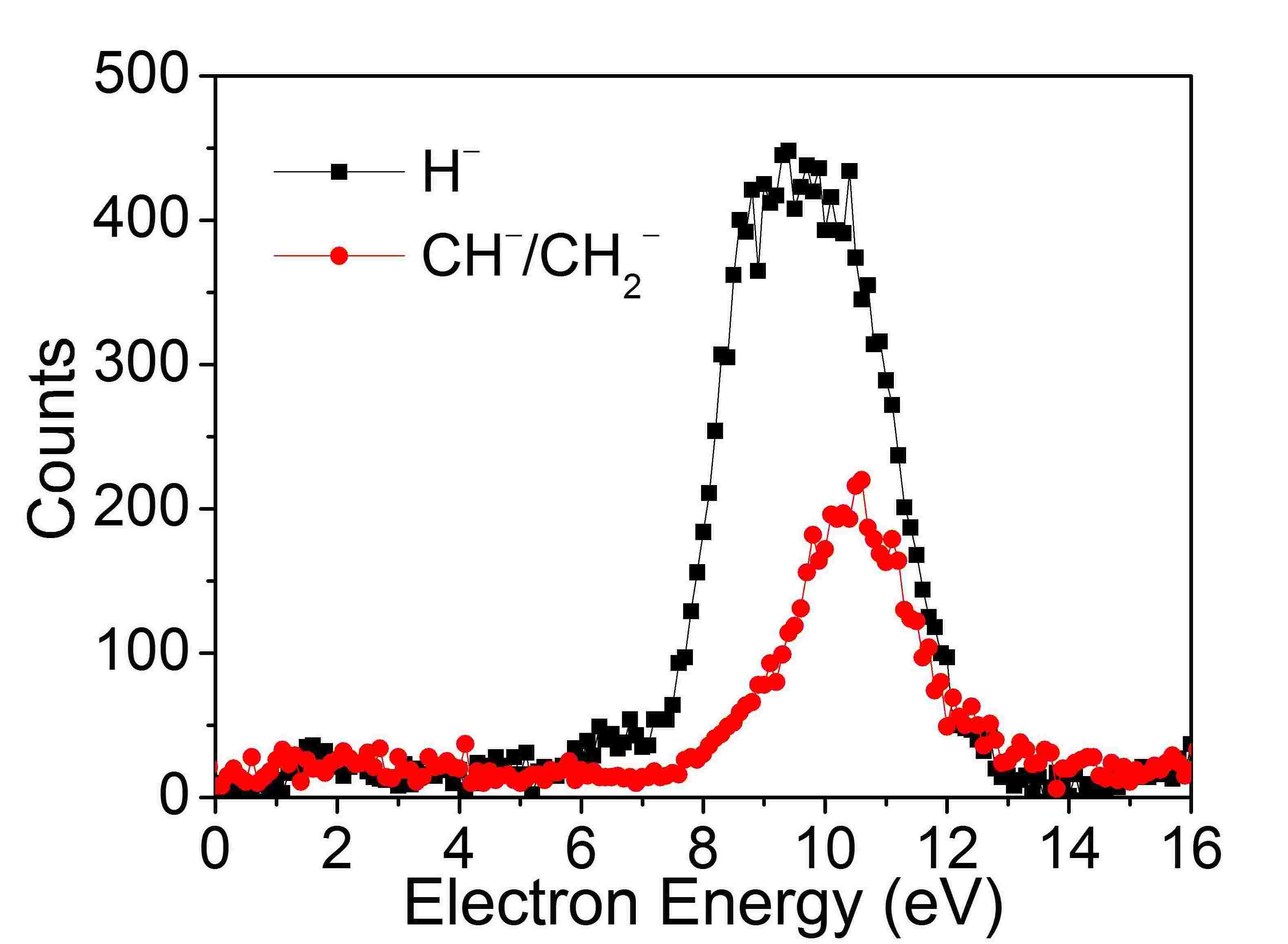}
\caption{Ion yield curve showing \ce{H-} and \ce{CH2^{-}}  ions produced from DEA to Methane. (not to scale) }
\label{fig5.11}
\end{figure}

Methane is another important molecule with its ubiquitous presence in planetary atmospheres, interstellar medium and many industrial applications. It is also crucial for applications involving plasma deposition of diamond like carbon films, diamond films and preparation of carbon nanotubes apart from its role in edge plasmas for fusion devices \cite{c5morgan}. While there exist electron impact, VUV absorption and photo-dissociation studies, there isn't a single report on dissociation dynamics due to low energy electron attachment. In one of the earliest experiments, Smith \cite{c5smith} reported observing \ce{H-}, \ce{C-}, \ce{CH-} and \ce{CH2^{-}} anion fragments formed by impact of 50 eV electrons on methane.  In 1963, Trepka and Neuert \cite{c5trepka} identified the negative ions resulting from DEA to methane to be \ce{H-} and \ce{CH2-} and obtained their ion yield curves. The only absolute cross section data are the total ion measurements by Sharp and Dowell \cite{c5sharpdowell2}. Only recently, the first ever measurements on absolute partial cross sections for the formation of \ce{H-} and \ce{CH2^{-}} from methane have been made by our lab recently \cite{c5rawat2}. Methane has a broad resonance peak centered at around 10 eV in the \ce{H-} channel, whereas the cross section for the heavier anion fragment of \ce{CH2^{-}} has a relatively narrow peak centered at 10.5 eV (see Figure \ref{fig5.11}). The cross sections for the \ce{CH2^{-}} channel are an order of magnitude smaller than those for the \ce{H-} channel \cite{c5rawat2}. The details of the kinetic energy and angular distribution measurements are given below.

\section{On the parent (\ce{CH4}) and grandparent (\ce{CH4^{+}}) electronic states of the resonance}

The ground state electronic configuration of methane is \ce{1a1^{2} 2a1^{2} 1t2^{6} -> 1A^{1}} state, where the highest occupied molecular orbital is a triply degenerate orbital (\ce{1t^{2}}). Optical absorption spectrum and electron energy loss experiments on methane \cite{c5robbins, c5harshbarger} are known to consist of two distinct bands centered at 78200 $cm^{-1}$ (9.7 eV) and 83600 $cm^{-1}$ (10.4 eV) due to \ce{1t2} to 3s Rydberg excitation. The spectrum of methane is mostly smooth devoid of any sharp bands/structures. This is due to the high symmetry of methane, which leads to complicated Jahn-Teller effects and due to the lack of lone pair orbitals in the molecule.  However, the photoelectron spectrum shows distinct structures while retaining similar envelope as the optical spectra. \ce{CH4^{+}} ion is found to have the lowest energy with \ce{D_{2d}} and \ce{C_{3v}} symmetry at the two bands. It is seen that the minimum energy geometry is reached by compressing the tetrahedron along an H-C-H bisector, tending to flatten the molecule (via e vibrations). At the minimum, the H-C-H angles are $94.78^{\circ}$ and $146.44^{\circ}$ with the CH distances larger than 0.01 A larger than in the ground state \cite{c5robbins}.

\section{Results and Discussion}

\begin{figure}[!h]
\centering
\subfloat[8 eV]{\includegraphics[width=0.3\columnwidth]{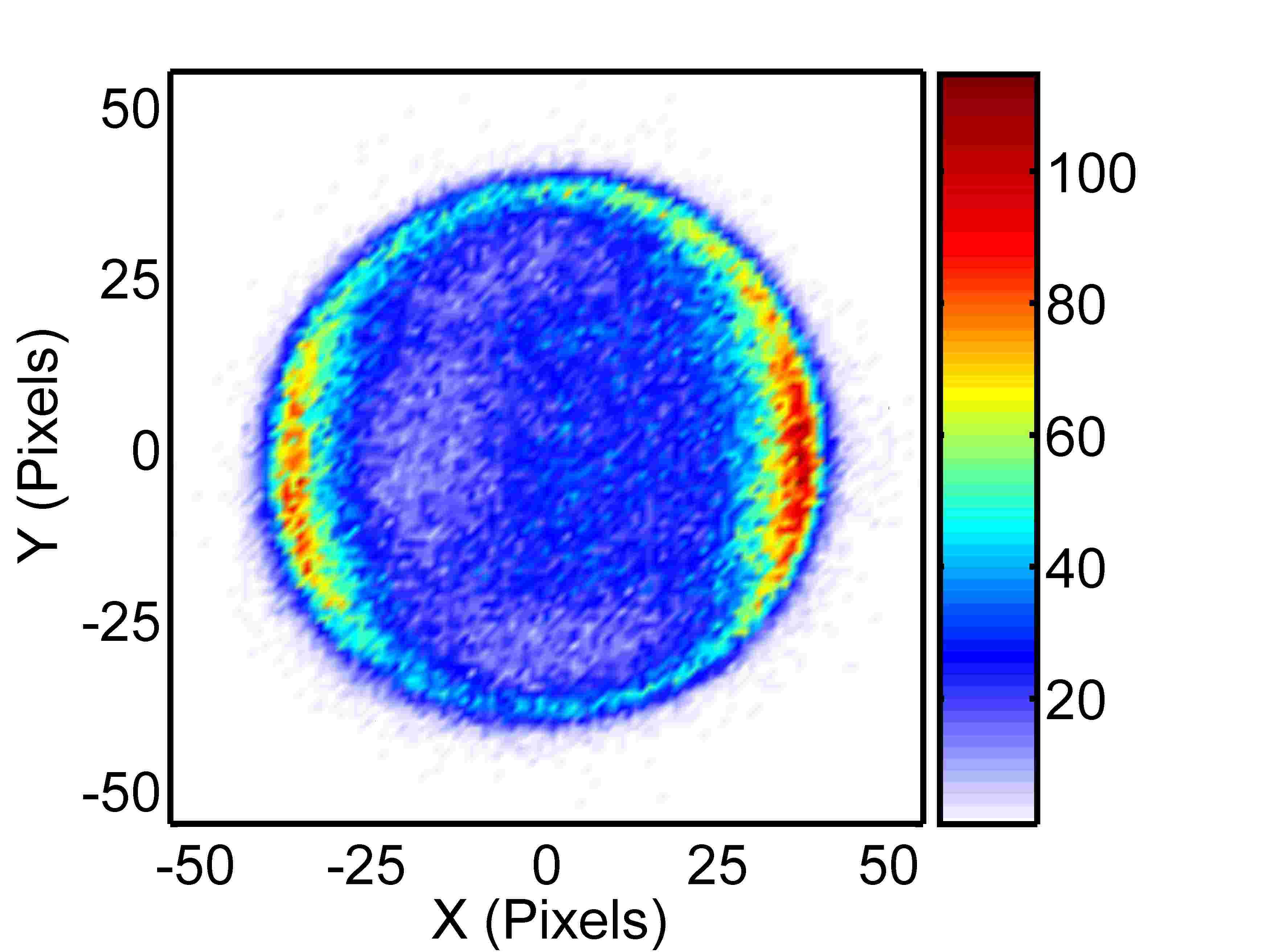}}
\subfloat[9 eV]{\includegraphics[width=0.3\columnwidth]{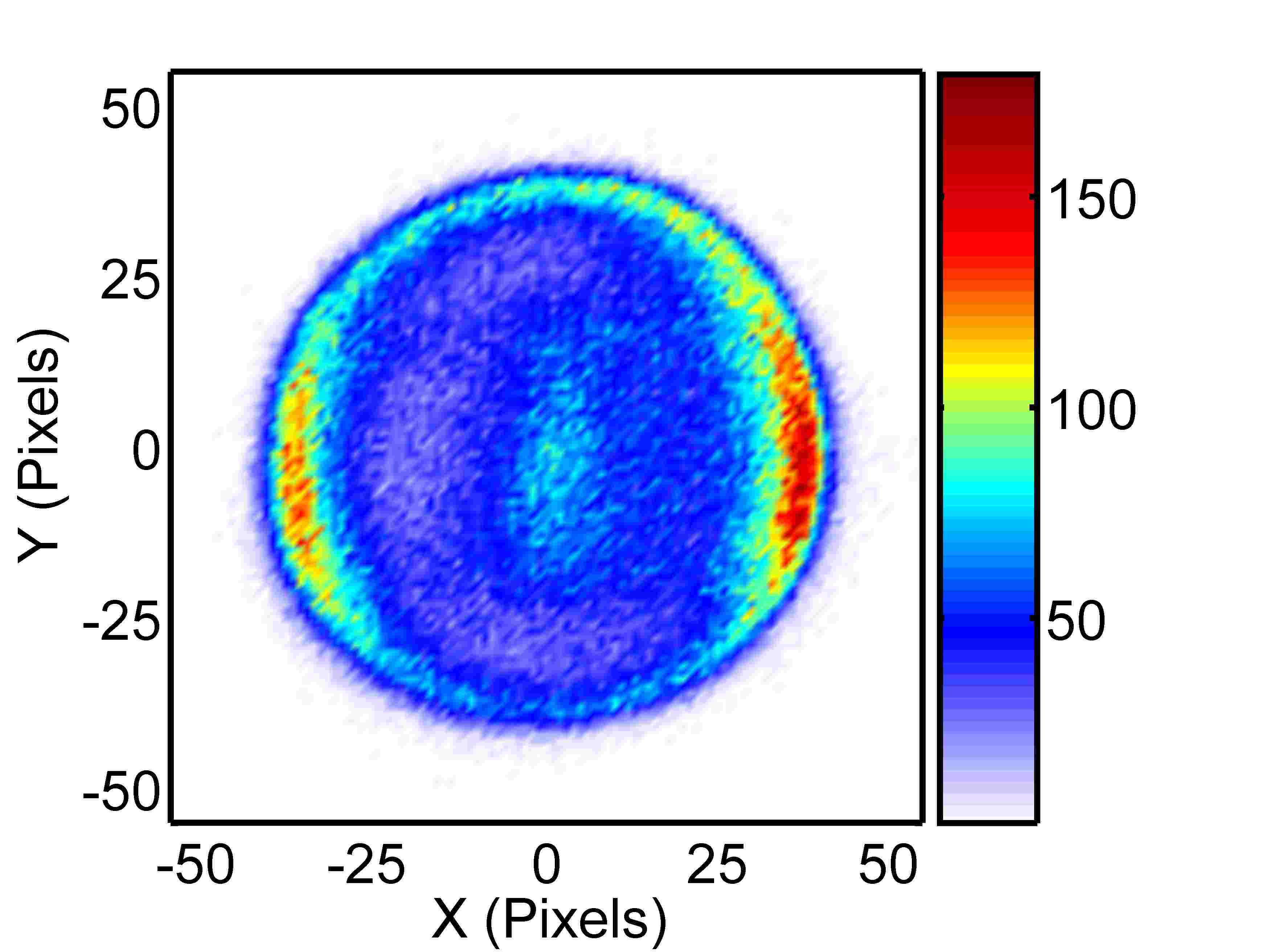}}
\subfloat[10 eV]{\includegraphics[width=0.3\columnwidth]{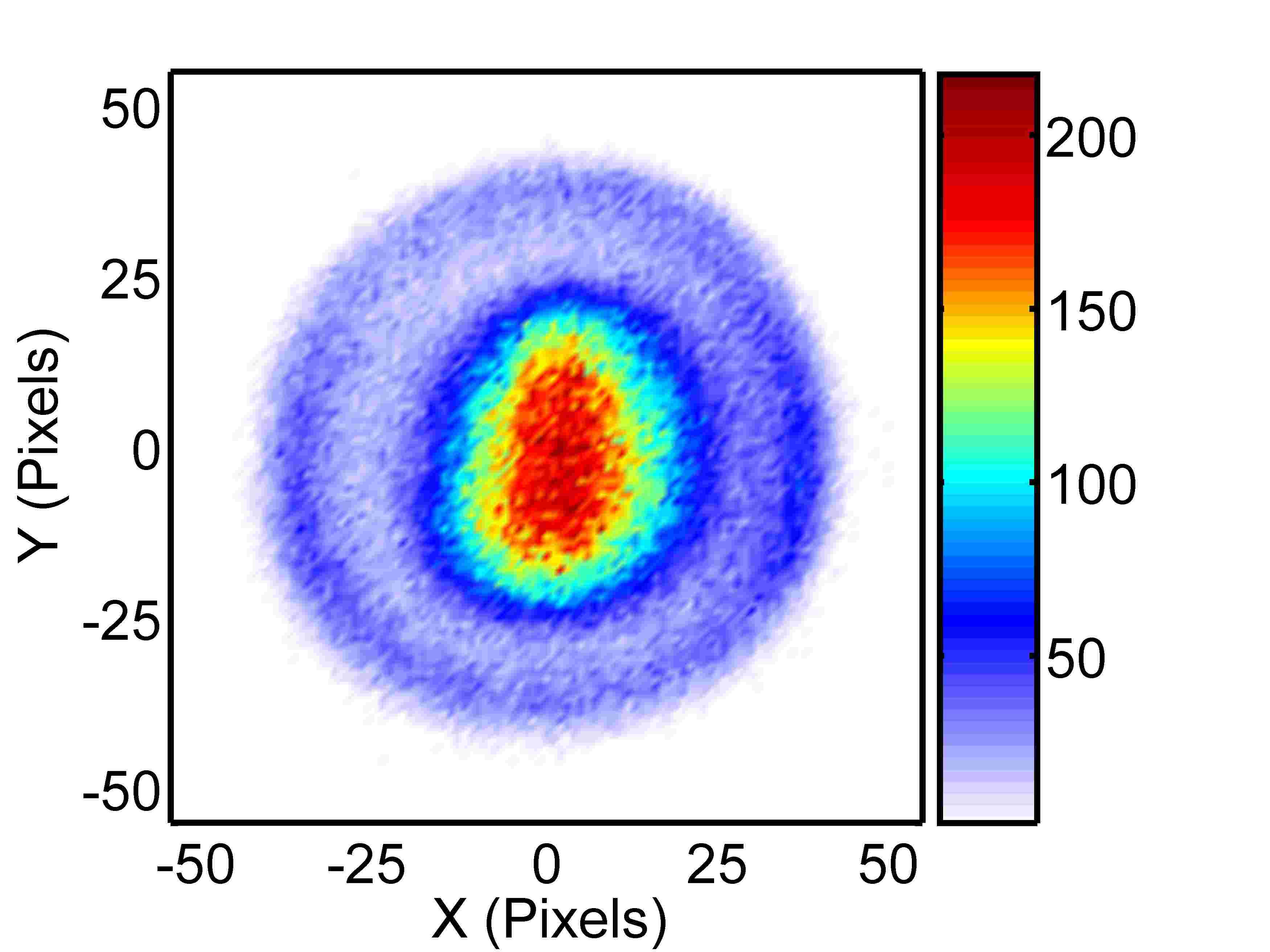}}\\
\subfloat[11 eV]{\includegraphics[width=0.3\columnwidth]{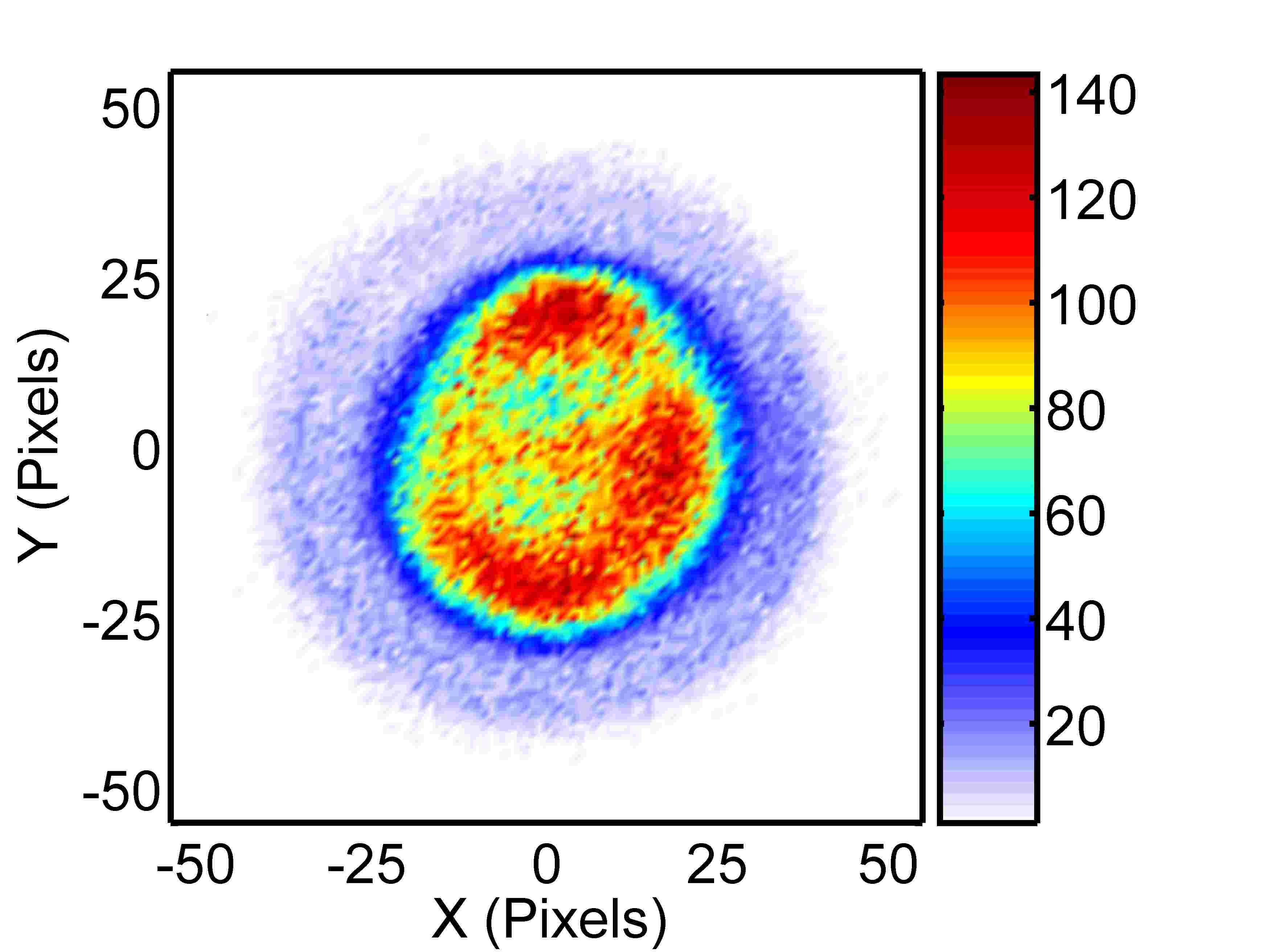}}
\subfloat[12 eV]{\includegraphics[width=0.3\columnwidth]{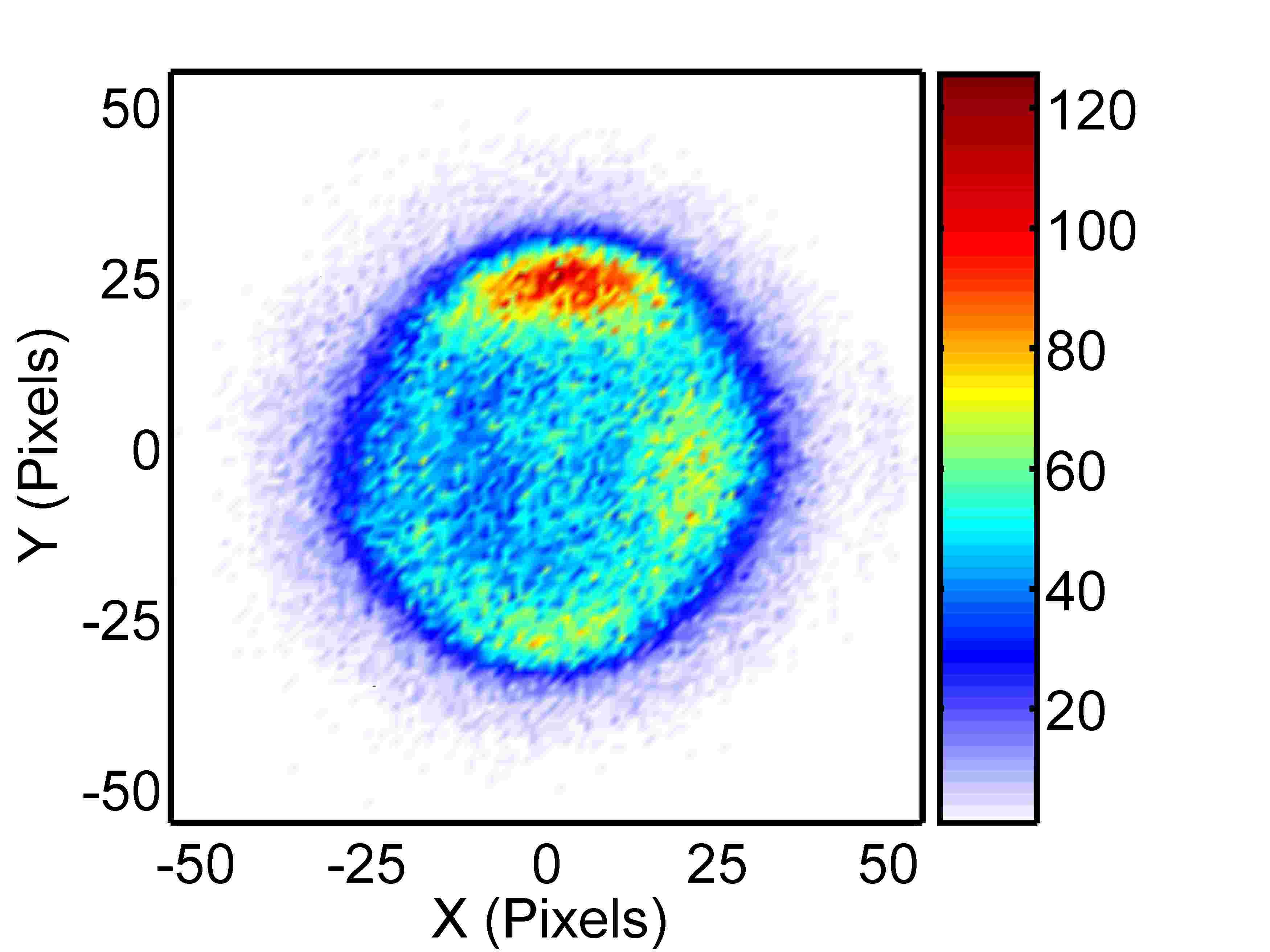}}
\caption{Velocity images of \ce{H-} ions from DEA to \ce{CH4}. The electron beam direction is from top to bottom in every image.}
\label{fig5.12}
\end{figure}

\begin{figure}[!ht]
\centering
\subfloat[9.5 eV]{\includegraphics[width=0.3\columnwidth]{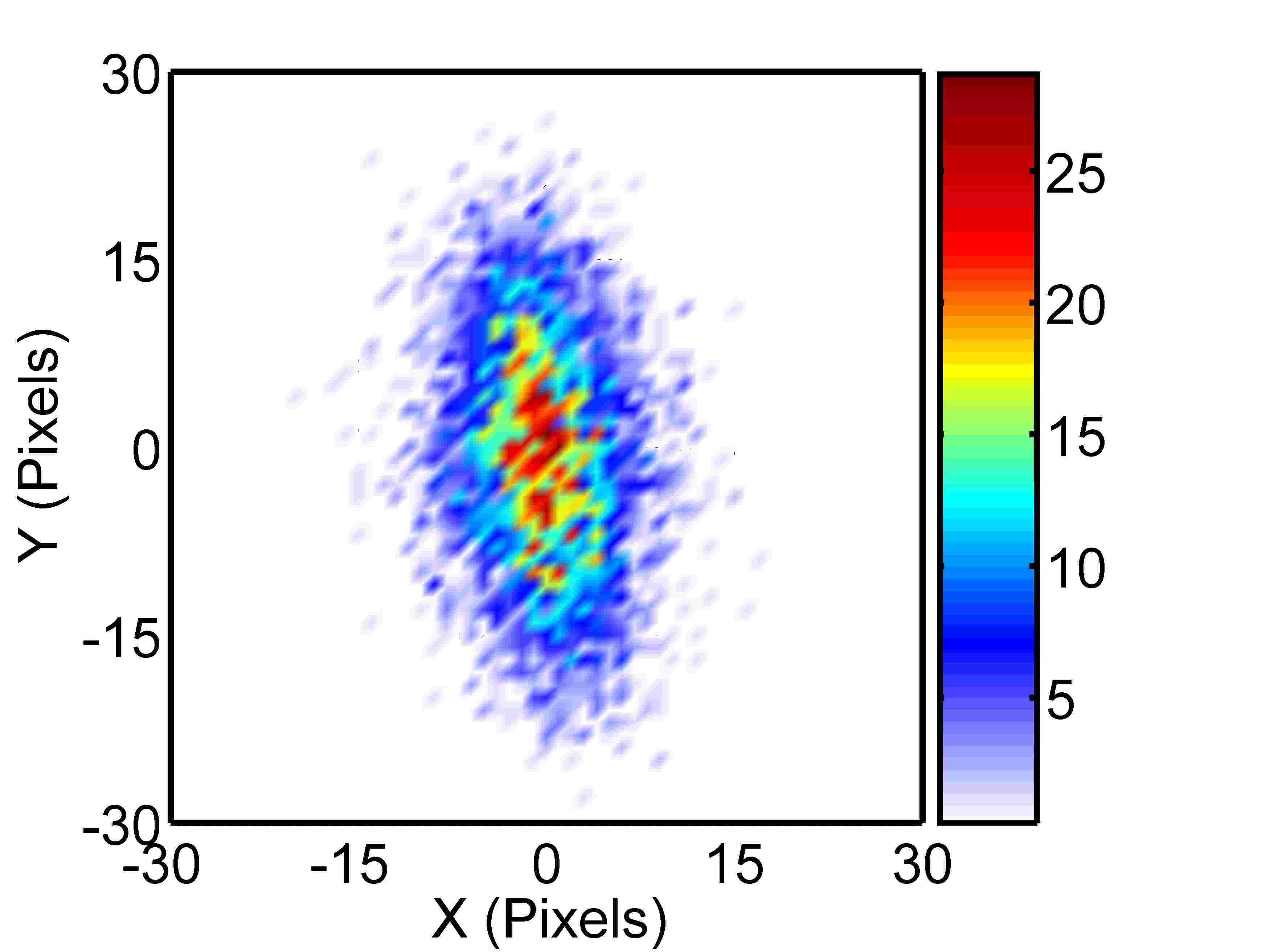}}
\subfloat[10.5 eV]{\includegraphics[width=0.3\columnwidth]{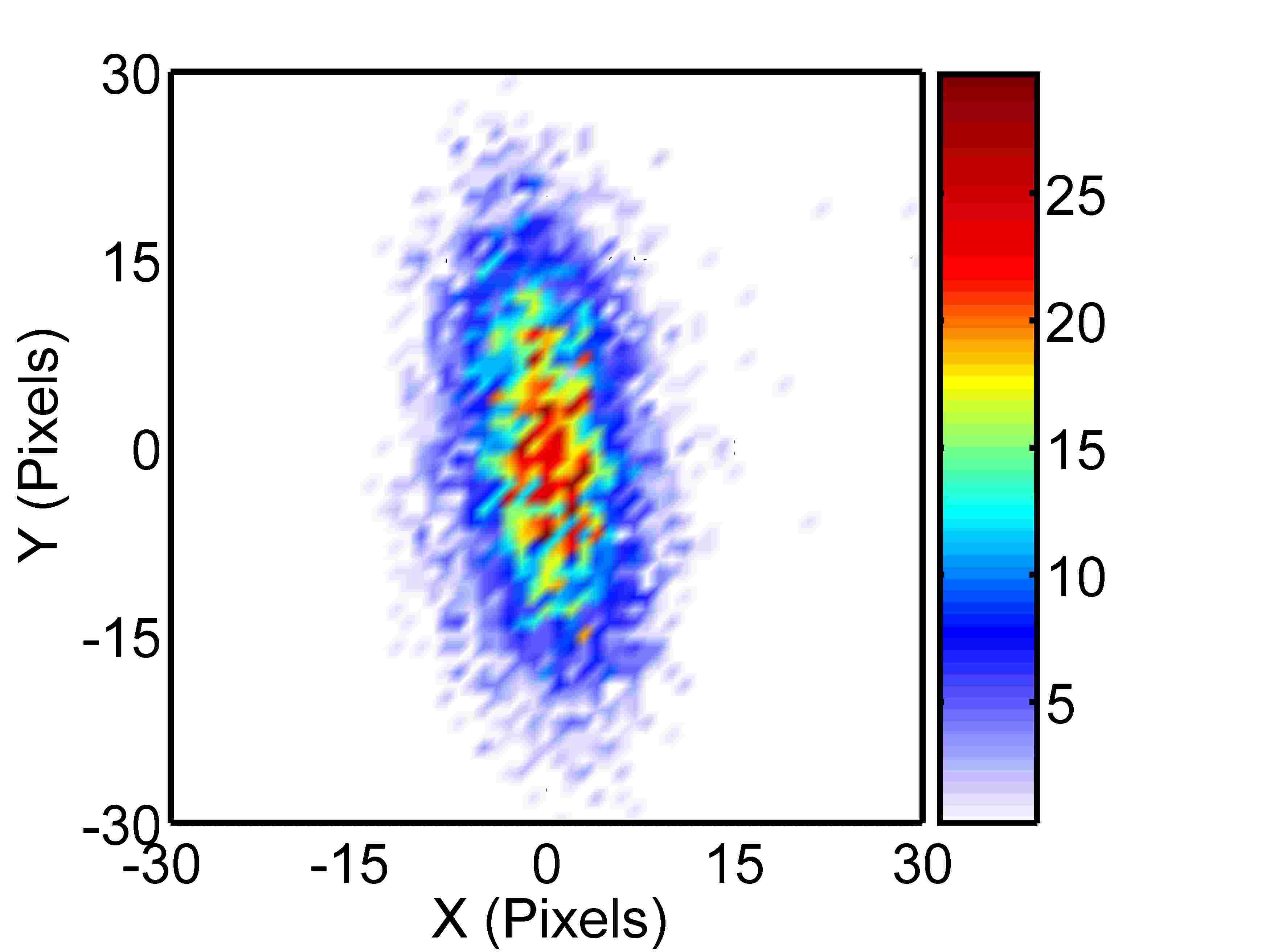}}
\subfloat[11.5 eV]{\includegraphics[width=0.3\columnwidth]{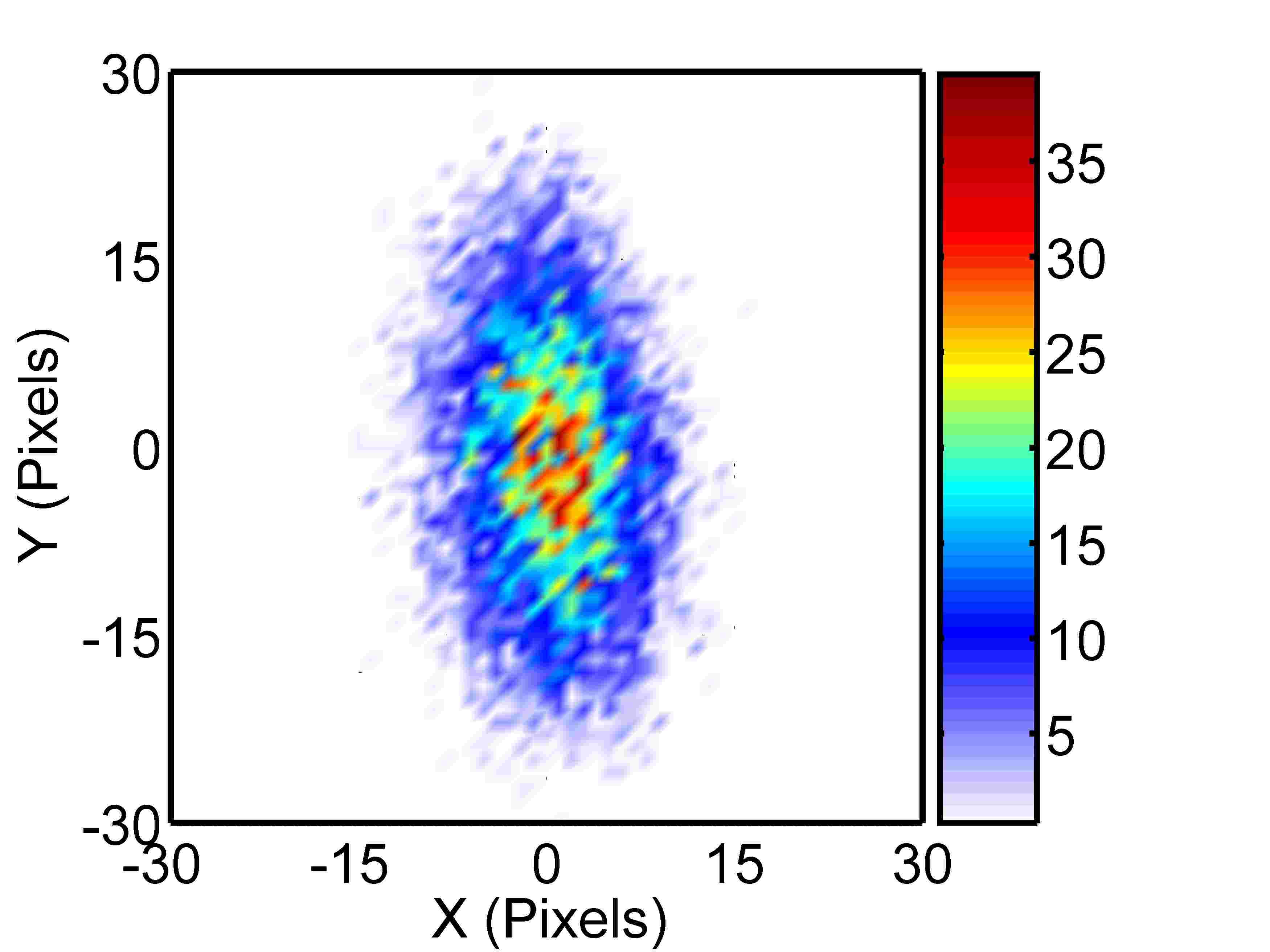}}
\caption{Velocity images of \ce{CH2^{-}} ions from DEA to \ce{CH4}. The electron beam direction is from top to bottom in every image.}
\label{fig5.13}
\end{figure}

The velocity images of \ce{H-} ions across the resonance are shown in Figure \ref{fig5.12} at incident electron energies 8, 9, 10, 11 and 12 eV respectively. Figure \ref{fig5.13} shows the velocity images of \ce{CH2^{-}}  fragment at 9.5, 10.5 and 11.5 eV respectively. From the images in Figure \ref{fig5.12}, it is seen that the scattering behaviour of \ce{H-} ions changes with increase in electron energy. At energies lower than 10 eV, we see an outer ring with maximum intensity in perpendicular direction to the electron beam and an unresolved inner structure. The outer ring fades away with increase in electron energy while the inner structure increases in intensity and evolves into a distinct pattern. These structures indicate the presence of more than one dissociation limit. Figure \ref{fig5.13} shows the images of the heavier anion fragment, \ce{CH2^{-}}. These are seen to have very little kinetic energy.

\begin{figure}[!htbp]
\centering
\subfloat[]{\includegraphics[width=0.45\columnwidth]{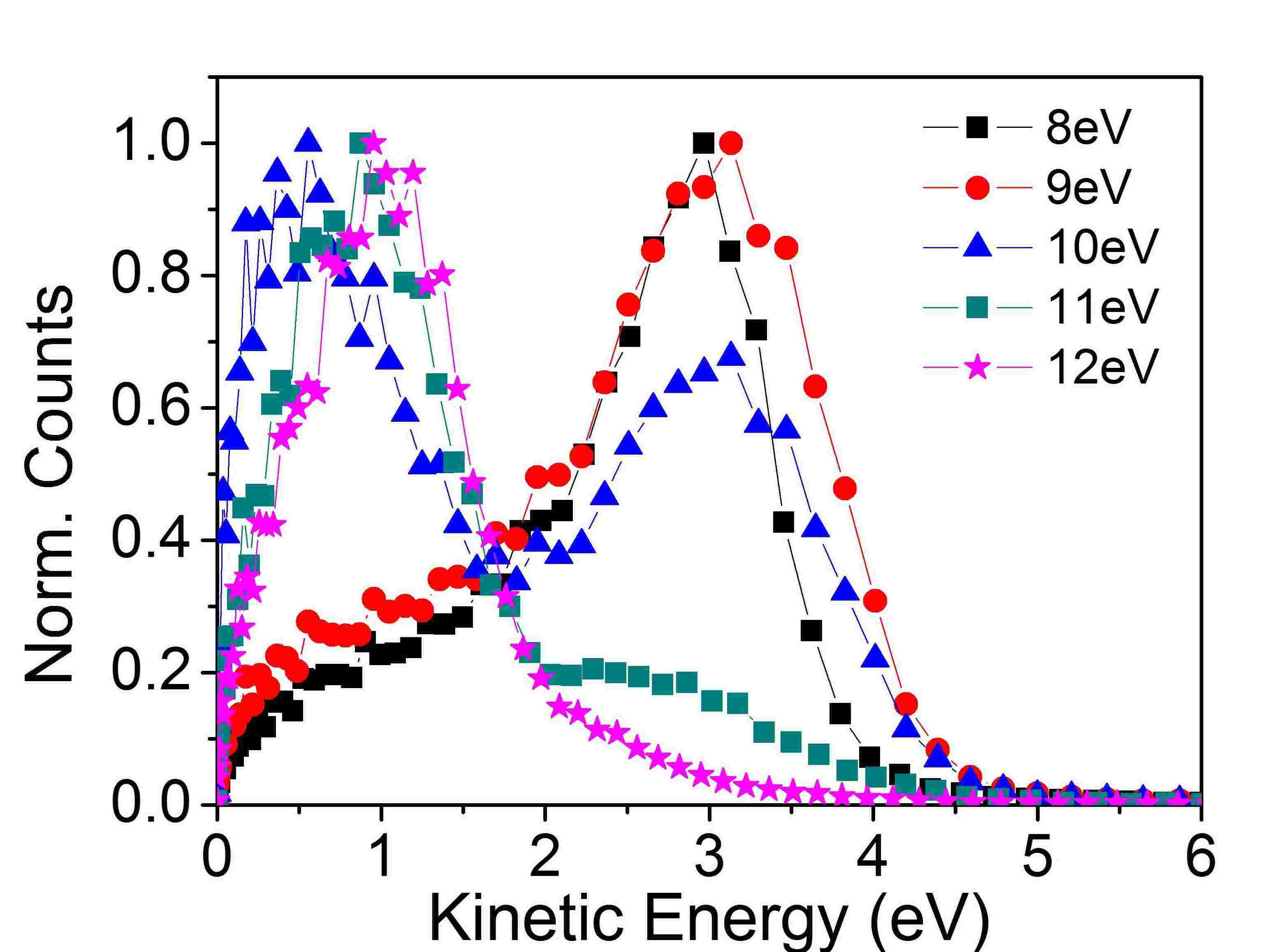}}
\subfloat[]{\includegraphics[width=0.45\columnwidth]{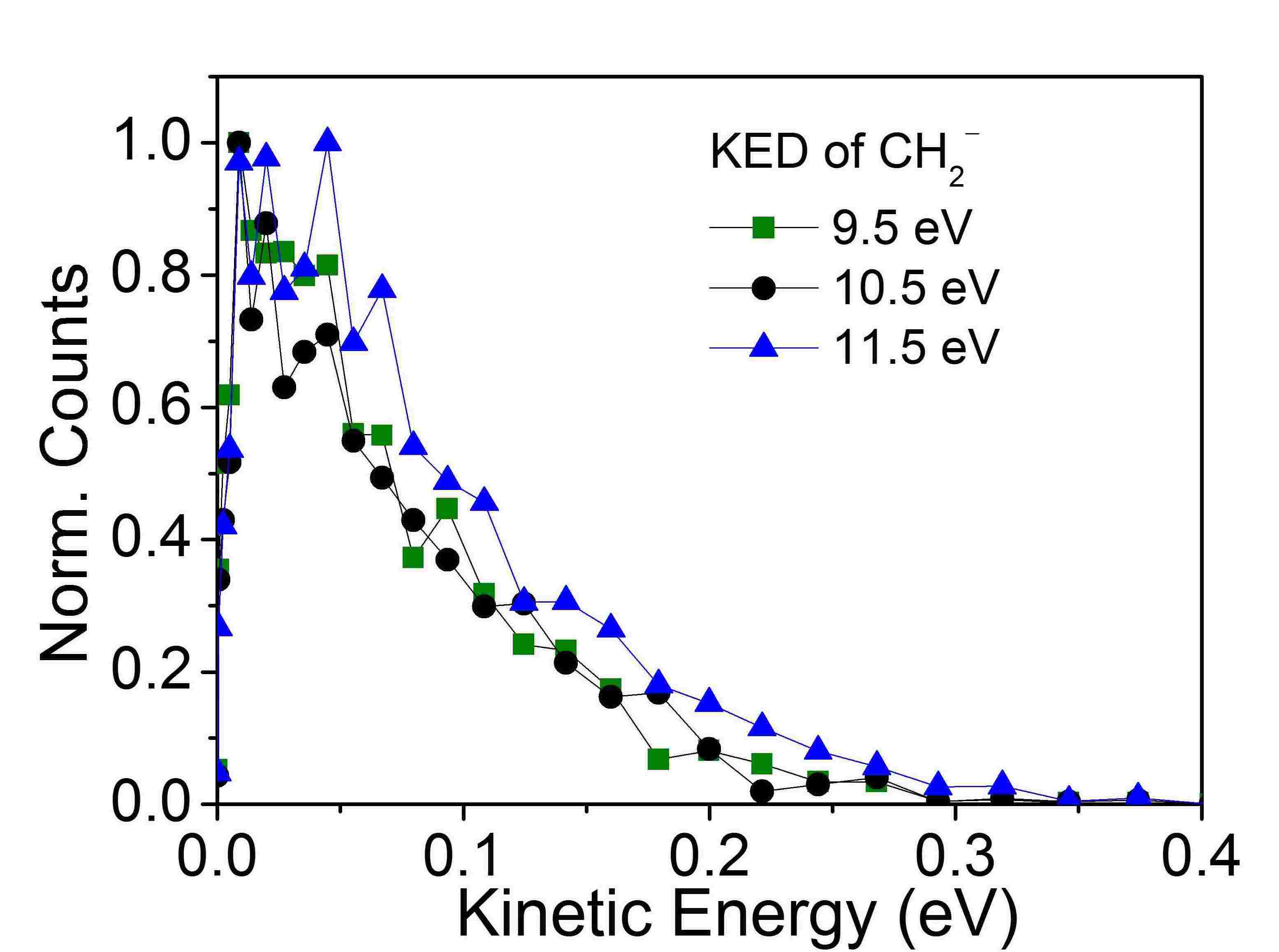}}
\caption{KE distribution of (a) \ce{H-} ions and (b) \ce{CH2^{-}} ions }
\label{fig5.14}
\end{figure}

To find out the dissociation channels leading to the formation of \ce{H-}, we look at the kinetic energy distribution (Figure \ref{fig5.14}) of \ce{H-} ions shown in the velocity images above. As expected, the two structures seen in the images yield two structures in the kinetic energy distribution - first one, between 0 to 2 eV and the second one between 2 to 4.5 eV with a peak at 3 eV. 

The various dissociation channels due to electron attachment to methane are listed below along with their appearance energy.

\begin{table}
\caption{Dissociation channels of \ce{CH4^{-*}} anion and threshold energy}
\begin{center}
\begin{tabular}{cc}
\hline
\\
Dissociation Channel & Threshold \\
\\
\hline
\\
\\
\ce{H- + CH3 (X ^{2}A2^{"})} & 3.75 eV\\
\\
\ce{H- + CH3^{*} ($\tilde{B}$ ^{2}A1^{'})} & 9.51 eV \\
\\
\ce{H- + H + CH2(^{1}A1)} & 8.37 eV \\
\\
\ce{H- + H2 + CH(^{2}$\Pi$)} & 8.5 eV \\
\\
\ce{H- + H + H + CH(^{2}$\Pi$)} & 13 eV \\
\\
\ce{CH2^{-} + H2} & 3.96 eV \\
\\
\ce{CH2^{-} + H + H} & 8.5 eV \\
\\
\hline
\end{tabular}
\end{center}
\label{tab1}
\end{table}

Amongst the channels mentioned in Table \ref{tab1}, only the first channel can give rise to \ce{H-} ions of KE 4.5 eV. To elaborate, given incident electron energy of 8 eV, the \ce{H- + CH3} channel has 8-3.75 = 4.25 eV as excess energy left in them. Assuming that all of this energy appears as translational KE of the fragments, the \ce{H-} takes away 15/16th of the excess energy i.e. approximately 4 eV, which is what we observe. At 9 and 10 eV, the maximum KE estimated are 4.9 and 5.9 eV, however we don't see the KE distribution extending beyond 4.5 eV. This indicates that the excess energy goes into internal excitation of the \ce{CH3} fragment. Thus, the structure in the KE spectra between 2 to 4.5 eV is attributed to the \ce{H- + CH3}(threshold: 3.75 eV) channel with high internal excitation of the \ce{CH3} fragment.

The structure between 0 and 2 eV becomes prominent at electron energies above 10 eV with a clearer angular distribution pattern. The possible channels that could give rise to \ce{H-} ions with energies below 2 eV are (i) \ce{H- + CH3^{*} ($\tilde{B}$ ^{2}A1^{$\prime$})} (threshold 9.51 eV) (ii) \ce{H- + H + CH2} (threshold 8.37 eV) or (iii) \ce{H- + H2 + CH (^{2}$\Pi$)} (threshold 8.5 eV). Experiments on methyl radical to characterize the excited electronic states have shown that \ce{CH3^{*} (^{2}A1^{$\prime$})} state is known to undergo rapid predissociation via H atom tunnelling and hence is not stable \cite{c5settersten,c5westre}. Thus, even if the methyl radical were to be formed in the first electronic excited state, it would undergo dissociation thereby giving rise to \ce{H- + H + CH2}.  Therefore, we conclude that the \ce{H-} ions with KE below 2 eV arise via a three body breakup on electron attachment. This conclusion is supported by the photodissociation experiments \cite{c5park,c5mordaunt} where the \ce{CH2(^{1}A1) + 2H} and \ce{CH(^{2}$\Pi$)+ H2 + H} were seen along with \ce{H + CH3} and \ce{CH2 (^{1}A1) + H2 (^{1}$\Sigma$_{g}^{+})} channels. In an instantaneous fragmentation process of \ce{CH4^{-*}} to \ce{H- + H + CH2} on the capture of a 10 eV electron, the maximum kinetic energy of \ce{H-} would be close to 0.7 eV (i.e. 7/16th of excess energy). Sequential fragmentation schemes such as \ce{H + CH3^{-*} -> H + CH2 + H-} account for very low kinetic energies as the H atom in the first step takes away most of the excess energy. Therefore, the \ce{CH2 (^{1}A1)} channel can only account for energies below 0.7 eV. Energies above 0.7 eV can be attributed to \ce{H-} ions flying away from the \ce{CH3} fragment close to \ce{CH + H2} or \ce{CH2 + H} dissociation limit.

The identity of the heavier anion fragments cannot be established in our experiment as our spectrometer cannot resolve the \ce{CH3^{-}}, \ce{CH2^{-}} or \ce{CH-} masses. The cross section measurements indicated the heavier fragment to be \ce{CH2^{-}} \cite{c5rawat2,c5smith,c5trepka}. Maximum KE of \ce{CH2^{-}} estimated at 10 eV would be about 0.6 eV assuming that the other fragment \ce{H2 (^{1}$\Sigma$_{g})} to be in ground vibrational state. If the \ce{H2} fragment dissociates further into 2H atoms, the maximum KE of \ce{CH2^{-}} would be 0.19 eV. The observed KE are much lower suggesting that \ce{CH2^{-}} and \ce{H2} are formed with vibrational excitation. Results from photodissociation experiments \cite{c5park,c5mordaunt} also support the formation of \ce{CH2-}.
 
\section{Angular curves for \ce{T_{d}} point group}

Methane is a highly symmetric molecule with a tetrahedral geometry (\ce{T_{d}} group) and has 24 invariant operations i.e. Identity, rotations by $\pm  120^{\circ}$ about 8 \ce{C3} axes, rotations by $180^{\circ}$ about 3 \ce{C2} axes, 6 rotary reflections by $90^{\circ}$ and reflection in 6 mirror planes where each plane contains a H-C-H bond and bisects the other H-C-H bond. The irreducible representations (or symmetry states) under the \ce{T_{d}} point group are \ce{A1}, \ce{A2} (one dimensional), E (two dimensional) and \ce{T1} and \ce{T2} (three dimensional). The character table and the basis functions expressed in terms of the spherical harmonics \cite{c5koster,c5cacelli} for the \ce{T_{d}} symmetry group are given in Table \ref{tab2}. We calculate the transition amplitude $<Negative Ion state| Partial Wave | Initial Neutral State>$ by using the appropriate basis functions and partial wave expressions in the dissociation frame and take the square of the modulus of the transition amplitude to obtain the expression for intensity. The expressions corresponding to the scattering intensities are listed in Appendix A at the end of this paper. Figures \ref{fig5.15}(a), (b), (c) and (d) show the angular distribution curves obtained for \ce{H-} ions produced from methane negative ion of \ce{A1}, E, \ce{T1} and \ce{T2} symmetries respectively in a two body like breakup under axial recoil approximation assuming equilibrium bond angle of $109.5^{\circ}$ and for partial waves upto $l$=3. We see that there is no basis function for \ce{A2} state available using spherical harmonics upto $l$=3.  

\begin{table}
\caption{Character table and basis functions in terms of spherical harmonics upto $l$=3 for \ce{T_{d}} point group.}
\begin{center}
\begin{tabular}{ccccccc}
\hline
\\
 & I & 8\ce{C3} & 3\ce{C2} & 6\ce{S4} & 6\ce{$\sigma$_{d}} & Basis functions (upto $l$=3) \\
\\
\hline
\\
\ce{A1} & 1 & 1 & 1 & 1 & 1 & $Y_{0}^{0}$; $Y_{0}^{0}$ \\
\\
\ce{A2} & 1 & 1 & 1 & -1 & -1 & - \\
\\
\ce{E} & 2 & -1 & 2 & 0 & 0 & ($Y_{2}^{0}, Y_{2}^{2}$) \\
\\
\ce{T1} & 3 & 0 & -1 & 1 & -1 & ($\sqrt{\frac{3}{8}}Y_{3}^{3} + \sqrt{\frac{5}{8}}Y_{3}^{1}, Y_{3}^{2}, -\sqrt{\frac{3}{8}}Y_{3}^{-3} + \sqrt{\frac{5}{8}}Y_{3}^{-1}$) \\
\\
\ce{T2} & 3 & 0 & -1 & -1 & 1 & ($Y_{1}^{1}, Y_{1}^{0}, Y_{1}^{-1}$) \\
&&&&&& ($Y_{2}^{-1}, Y_{2}^{-2}, Y_{2}^{1}$) \\
&&&&&& ($\sqrt{\frac{5}{8}}Y_{3}^{3} - \sqrt{\frac{3}{8}}Y_{3}^{1}, Y_{3}^{0}, -\sqrt{\frac{5}{8}}Y_{3}^{-3} - \sqrt{\frac{3}{8}}Y_{3}^{-1}$) \\
\\
\hline
\end{tabular}
\end{center}
\label{tab2}
\end{table}

\begin{figure}[!htbp]
\centering
\subfloat[\ce{A1 -> A1}]{\includegraphics[width=0.4\columnwidth]{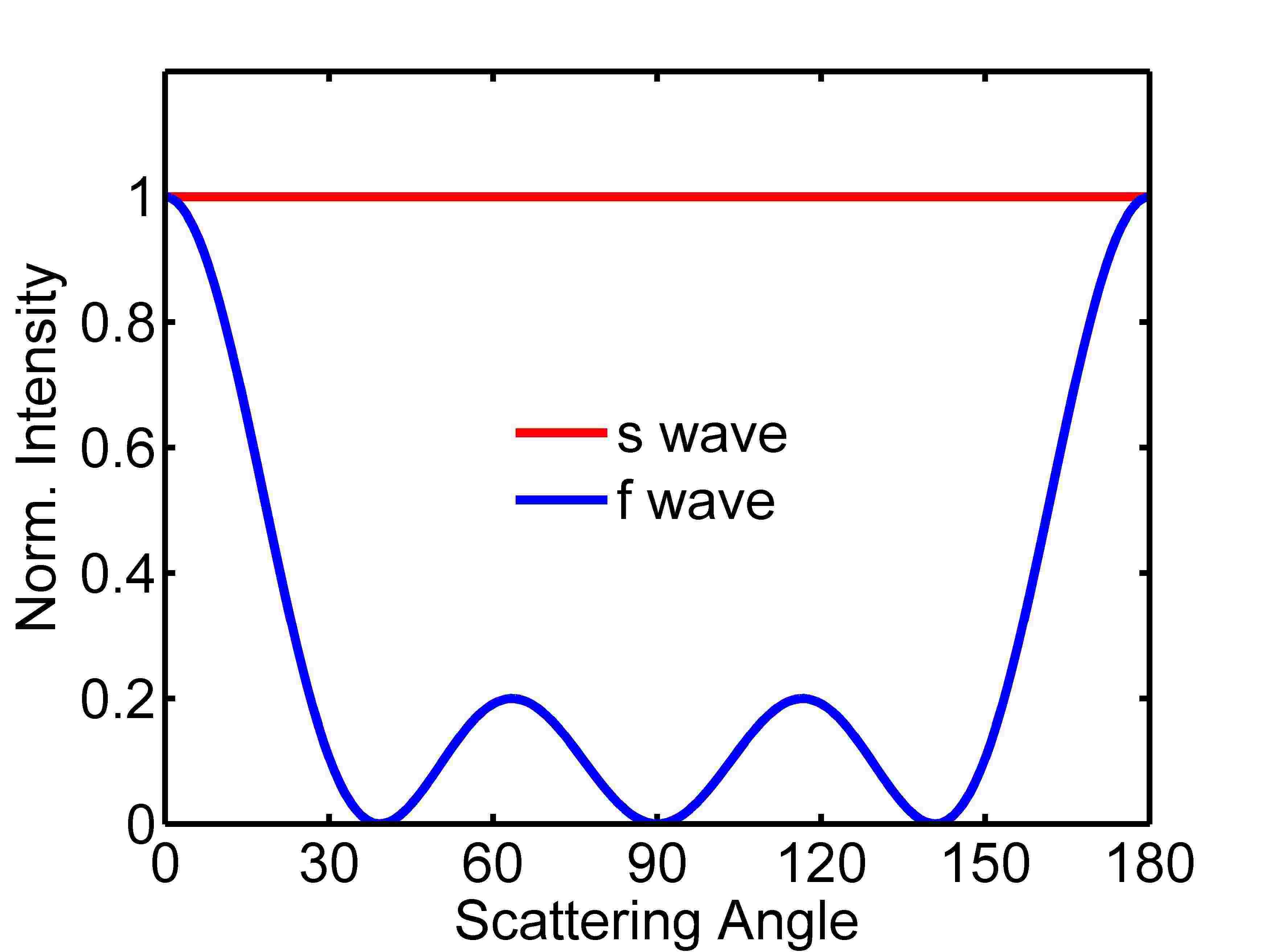}}
\subfloat[\ce{A1 -> E}]{\includegraphics[width=0.4\columnwidth]{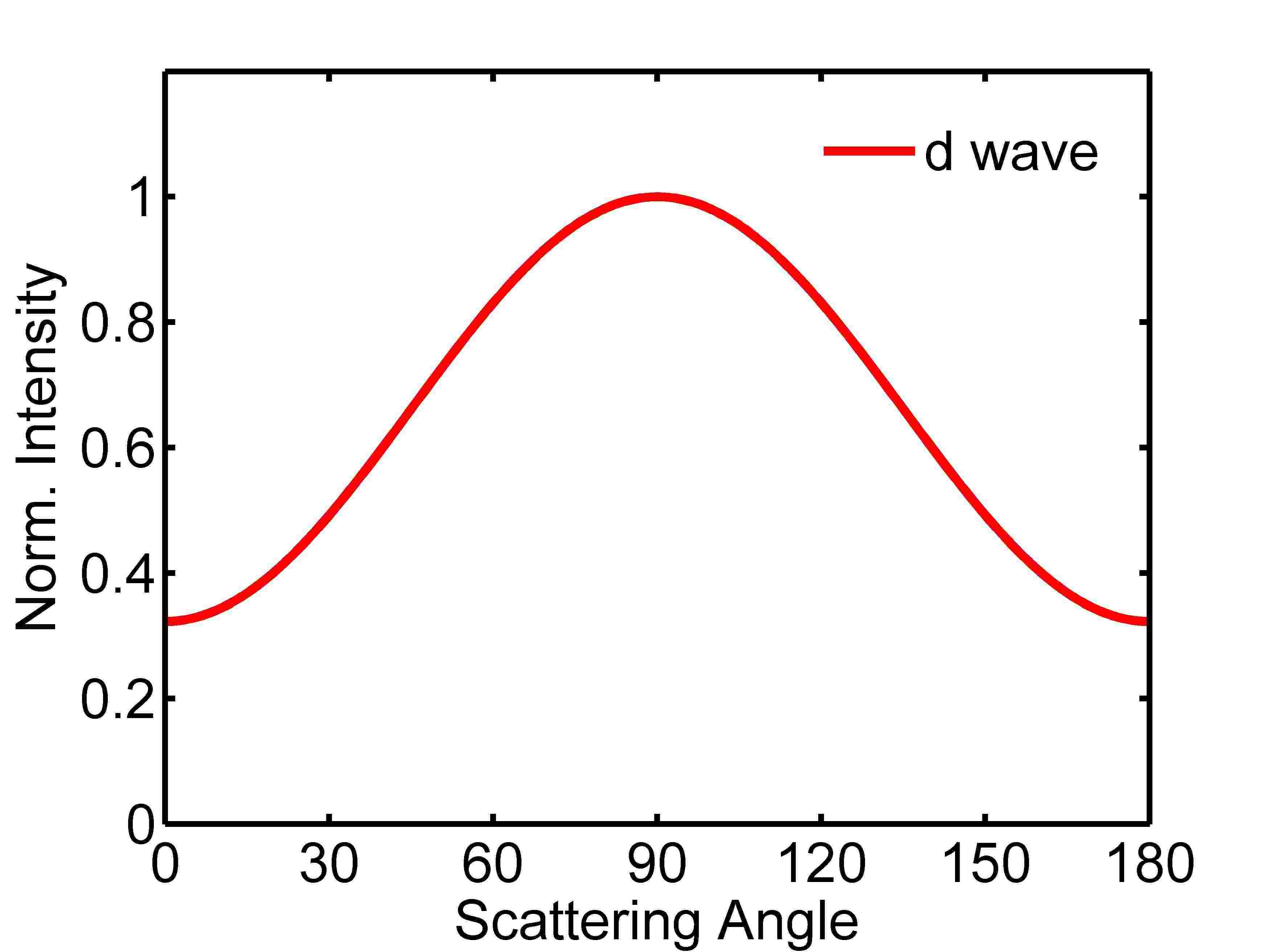}}\\
\subfloat[\ce{A1 -> T1}]{\includegraphics[width=0.4\columnwidth]{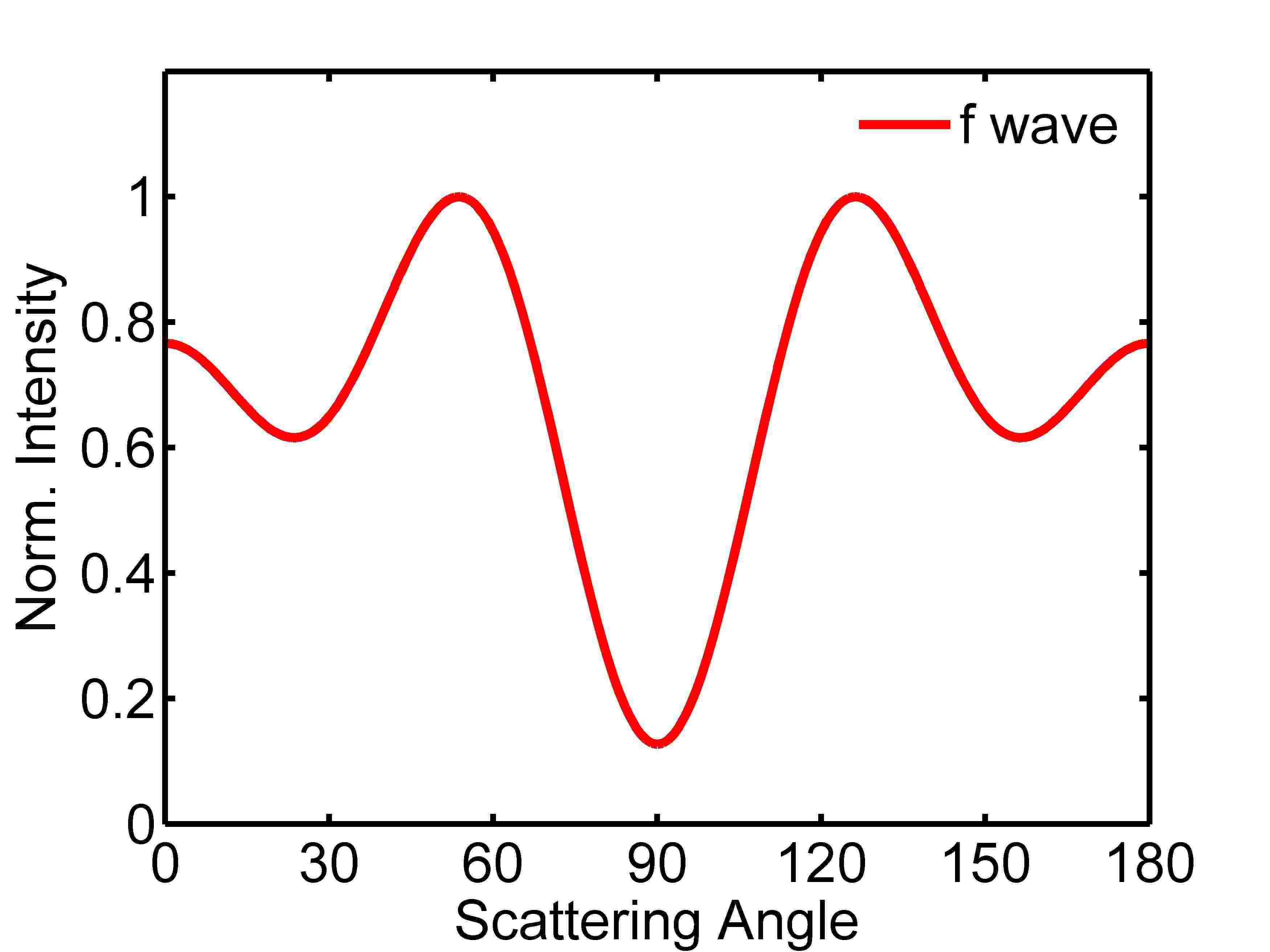}}
\subfloat[\ce{A1 -> T2}]{\includegraphics[width=0.4\columnwidth]{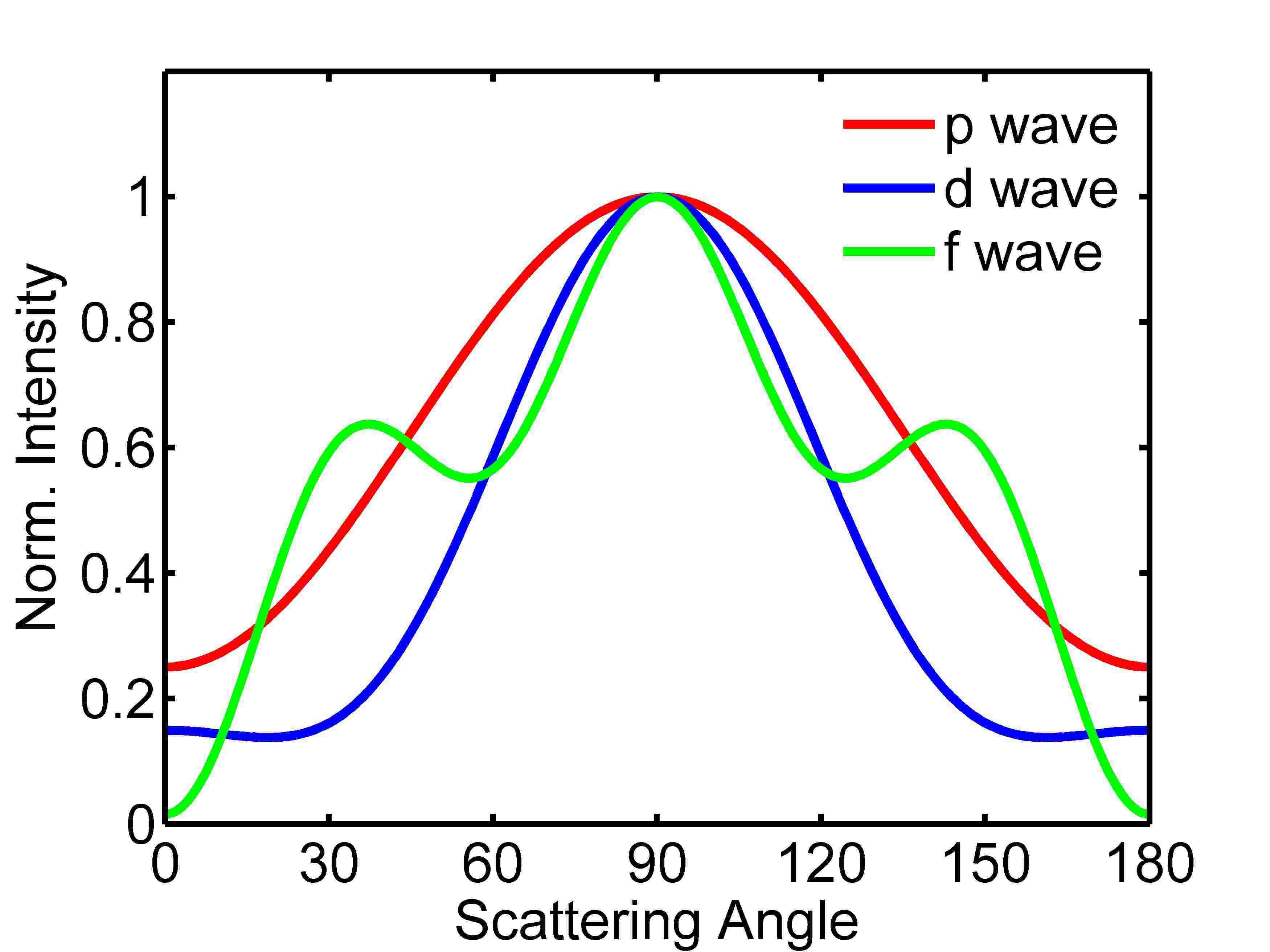}}
\caption{Angular curves for various symmetries under \ce{T_{d}} group.}
\label{fig5.15}
\end{figure}

\section{Angular distributions of \ce{H-} ions}
\begin{figure}[!htbp]
\centering
\subfloat[8 eV]{\includegraphics[width=0.3\columnwidth]{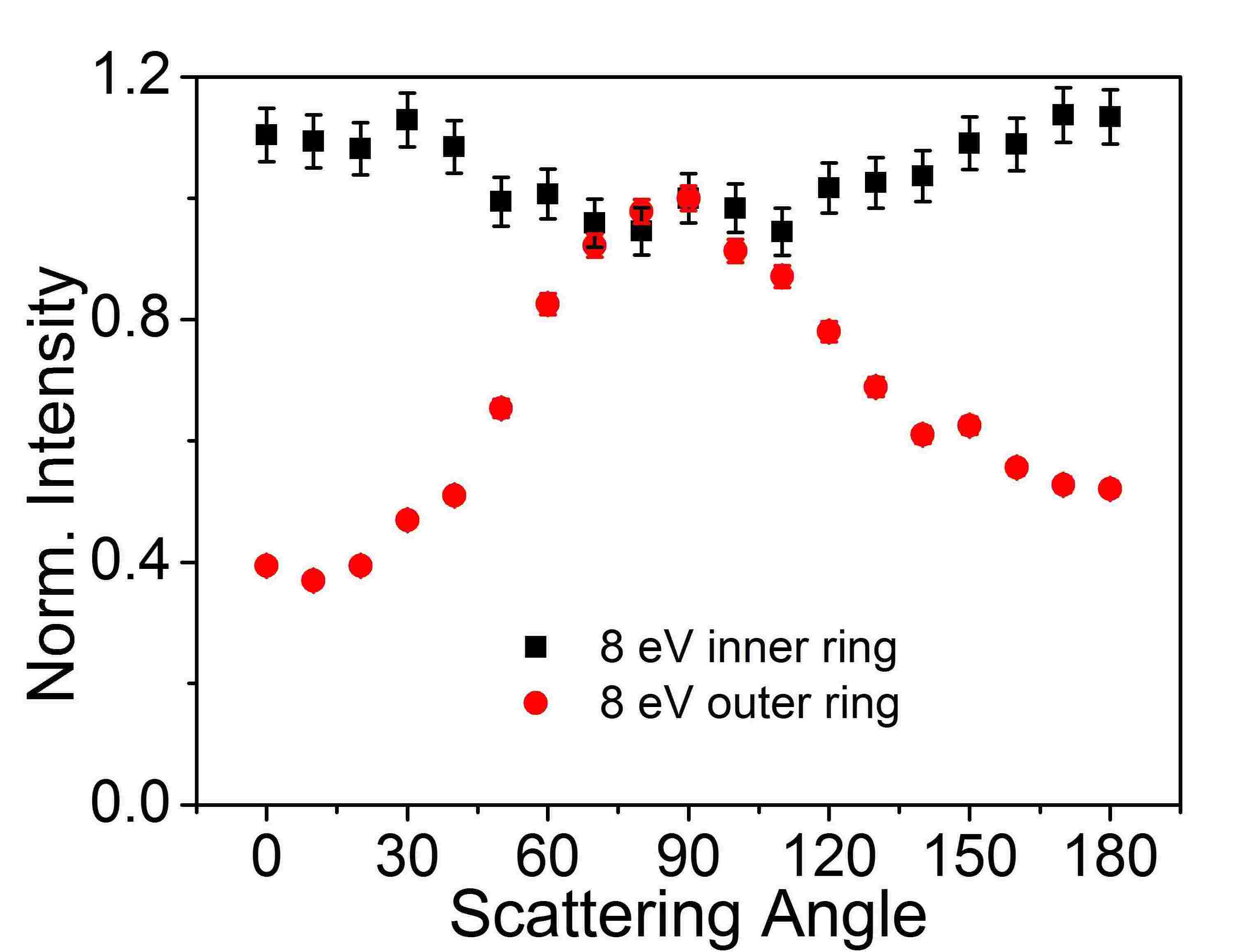}}
\subfloat[9 eV]{\includegraphics[width=0.3\columnwidth]{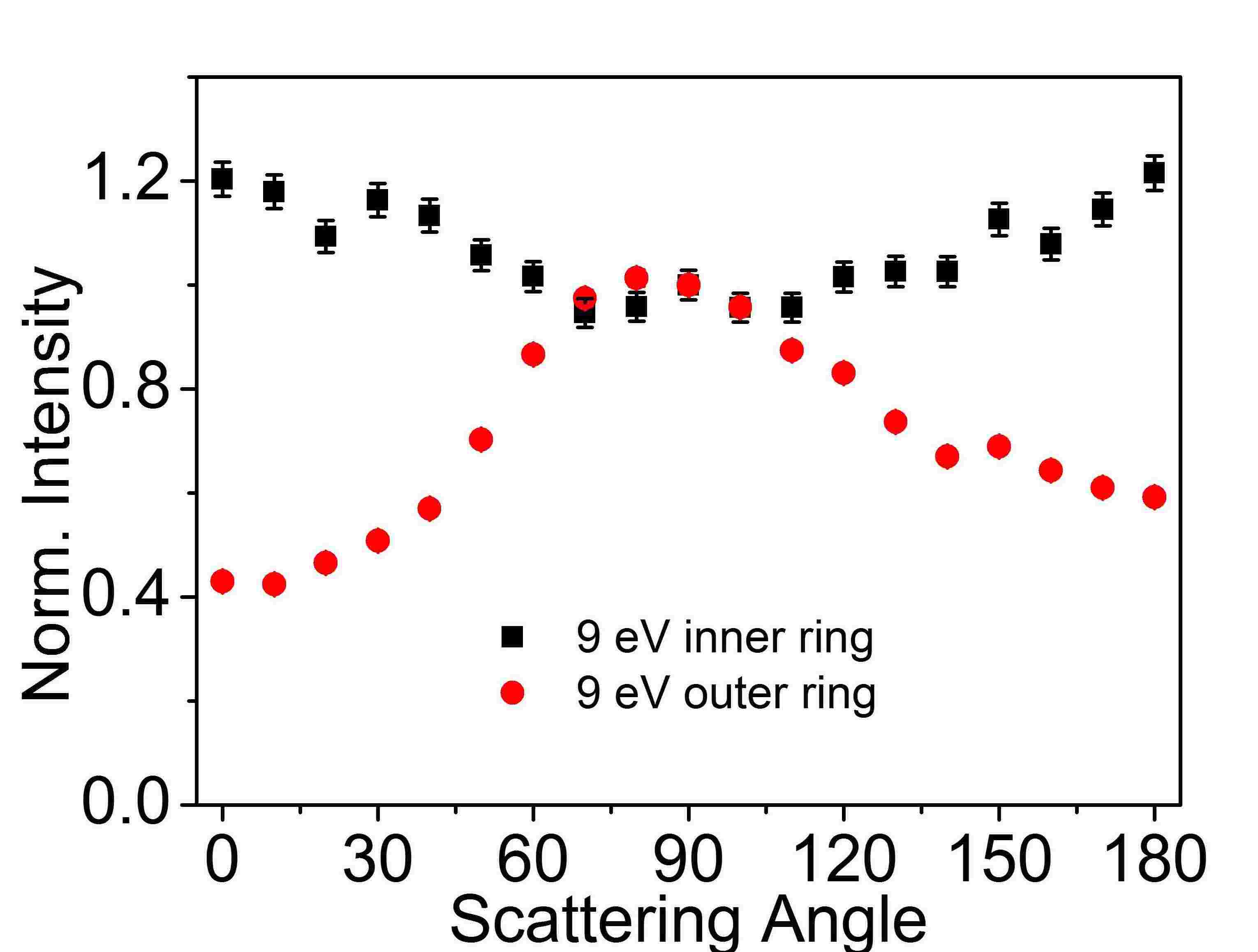}}
\subfloat[10 eV]{\includegraphics[width=0.3\columnwidth]{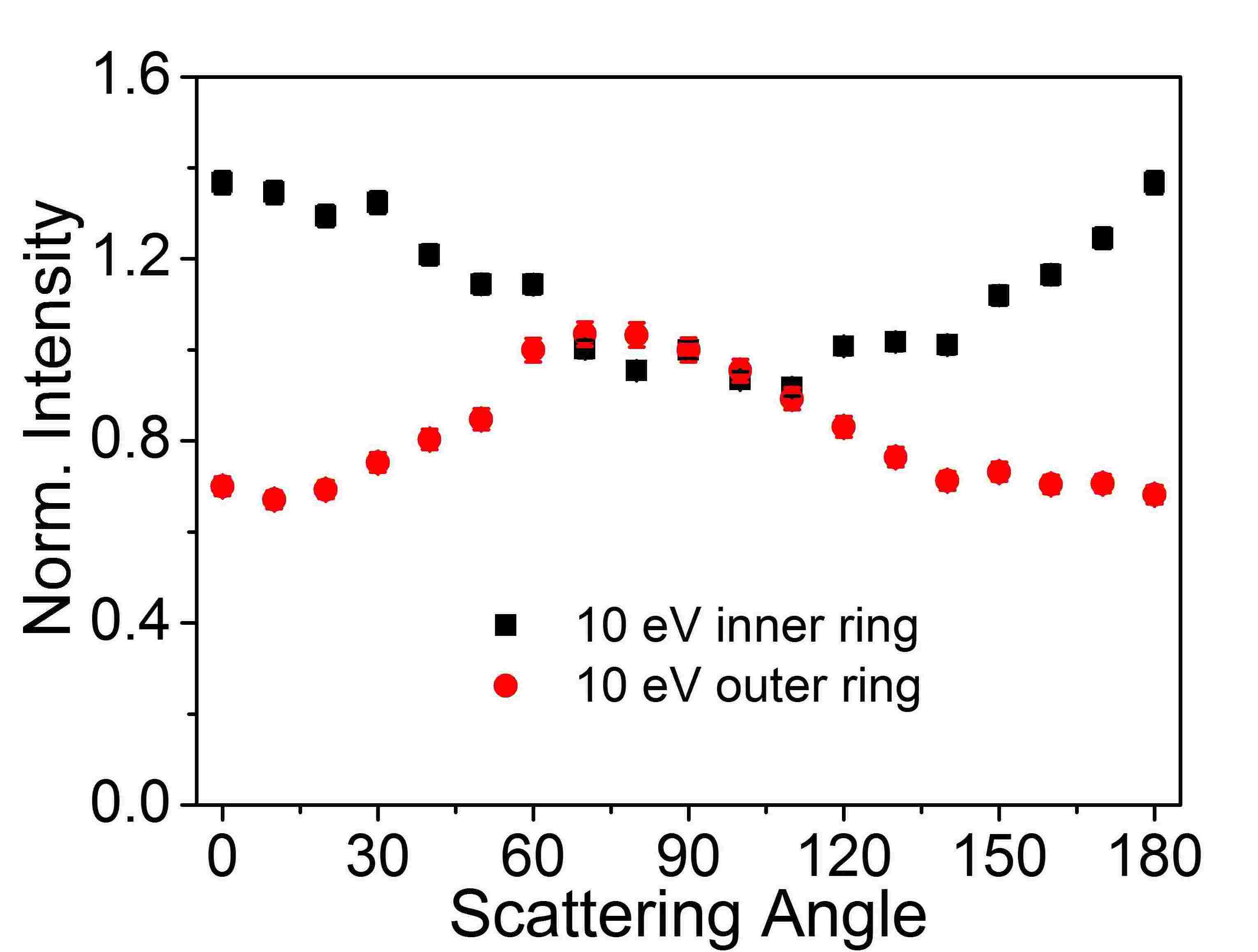}}\\
\subfloat[11 eV]{\includegraphics[width=0.3\columnwidth]{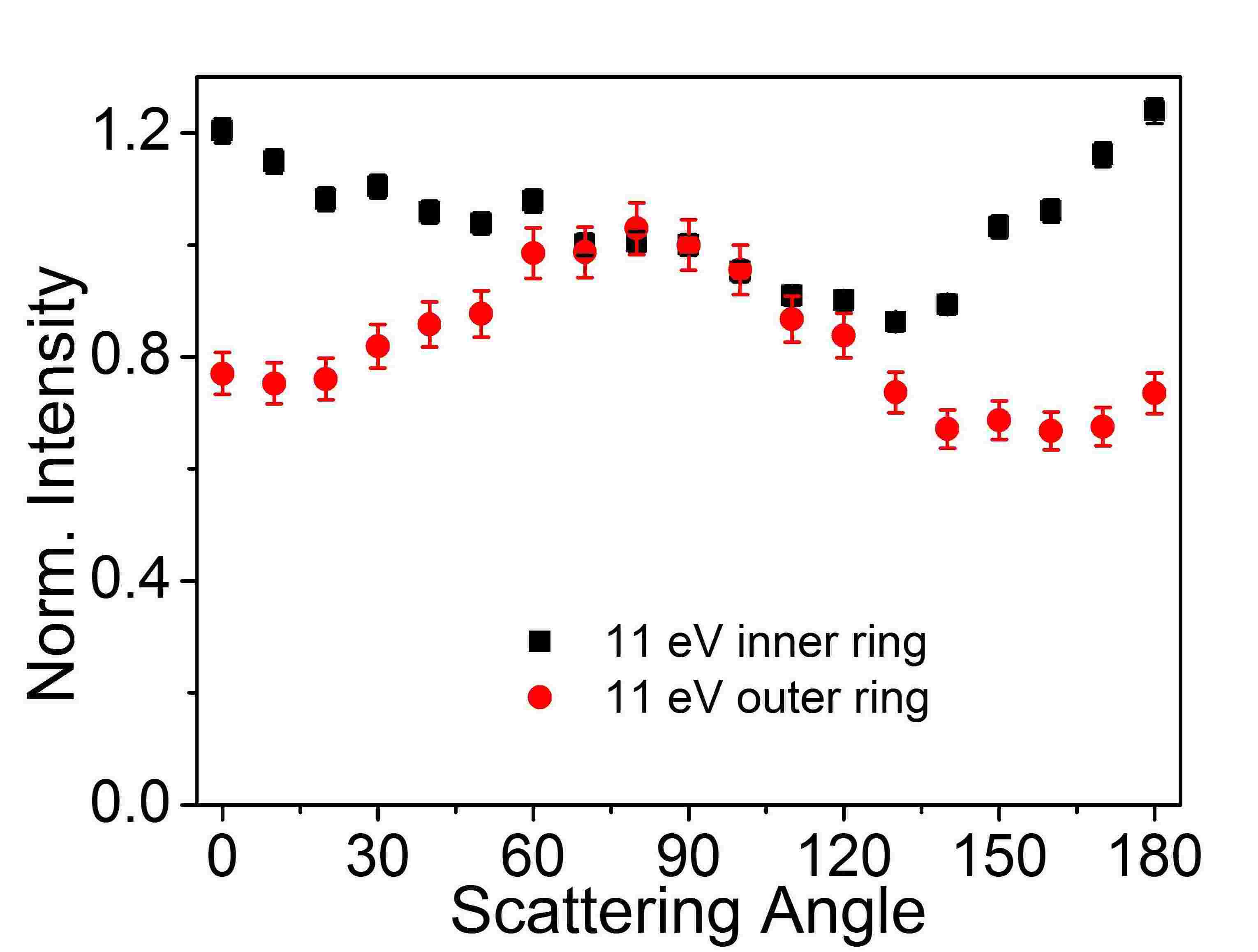}}
\subfloat[12 eV]{\includegraphics[width=0.3\columnwidth]{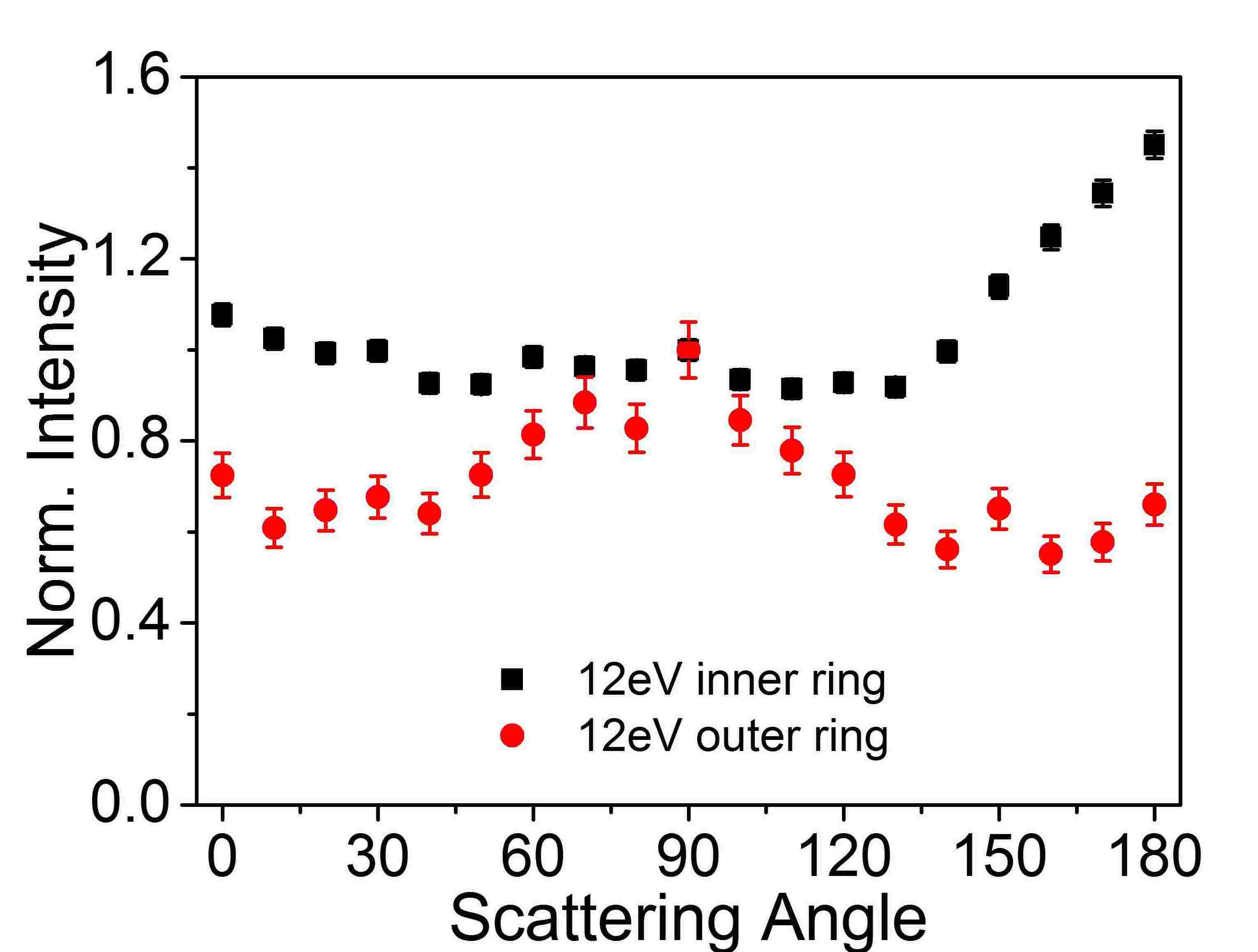}}
\caption{Angular distribution plots of \ce{H-} ions produced from the outer ring (red circles) and inner ring (black squares) at different incident electron energies.}
\label{fig5.16}
\end{figure}

From the velocity images of \ce{H-}, two distinct structures are seen; outer ring corresponding to the formation of \ce{CH3} radical in its electronic ground state and the inner ring corresponding to a three body fragmentation channel as deduced from the KE distribution. Figure \ref{fig5.16} plots the angular distribution of \ce{H-} ion in these two structures. The inner ring distribution contains ions from 0 to about 2 eV and the outer ring distribution contains ions above 2 eV. The outer ring has a broad peak about $90^{\circ}$ direction with sufficient intensity in the forward-backward angles. The inner ring has a forward-backward scattering distribution with a broad valley about $90^{\circ}$. At 12 eV, the backward angle becomes more intense than other angles.

The presence of two dissociation limits with different angular distribution as a function of electron energy suggests two closely spaced resonances seen as one broad resonance structure as in Figure \ref{fig5.11}. This leads us to the question of identifying the parent electronic states of the observed \ce{CH4^{-*}} resonance.  Looking at the hierarchy of molecular orbitals in the neutral \ce{CH4}, the highest occupied molecular orbital (HOMO) is a triply degenerate \ce{1t2} orbital and the next vacant orbital is \ce{3a1}. Presence of the triply degenerate orbital is known to cause Jahn-Teller distortion. The Jahn-Teller theorem states that any non-linear molecule with a degenerate electronic ground state will undergo a geometrical distortion that removes the degeneracy, because the distortion lowers the overall energy of the complex. Thus, electronic excitation of the triply degenerate \ce{1t2} orbital lowers the symmetry of the molecule and splits the \ce{t2} orbital into three non-degenerate states. The two dissociation limits may arise from non-degenerate orbitals of two different symmetries split up following the excitation of the triply degenerate \ce{t2} orbital under \ce{T_{d}} group. 

It could be the case that the second resonance is caused by the excitation of the penultimate HOMO-1 molecular orbital (\ce{2a1}). However, VUV absorption studies and ab initio calculations \cite{c5lee,c5mebel} have shown that the broad absorption feature from 9 to 12 eV is due to transition to a single triply degenerate excited state and has no \ce{2a1} contribution. Thus, in all likelihood the broad resonance structure of \ce{CH4^{-*}} in our DEA experiment comes from the excitation of the degenerate \ce{1t2} orbital only.

\begin{figure}[!htbp]
\centering
\subfloat[8 eV]{\includegraphics[width=0.3\columnwidth]{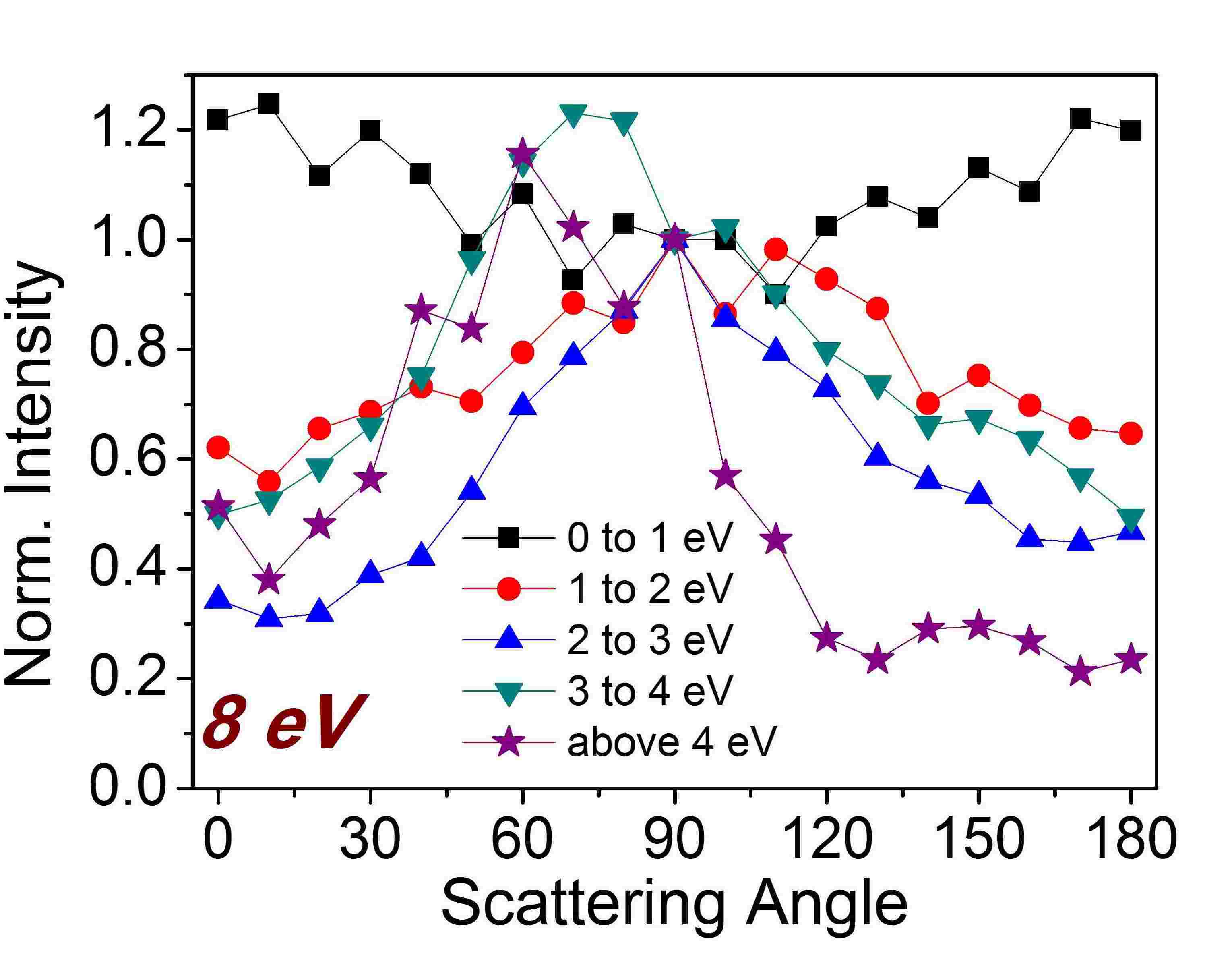}}
\subfloat[9 eV]{\includegraphics[width=0.3\columnwidth]{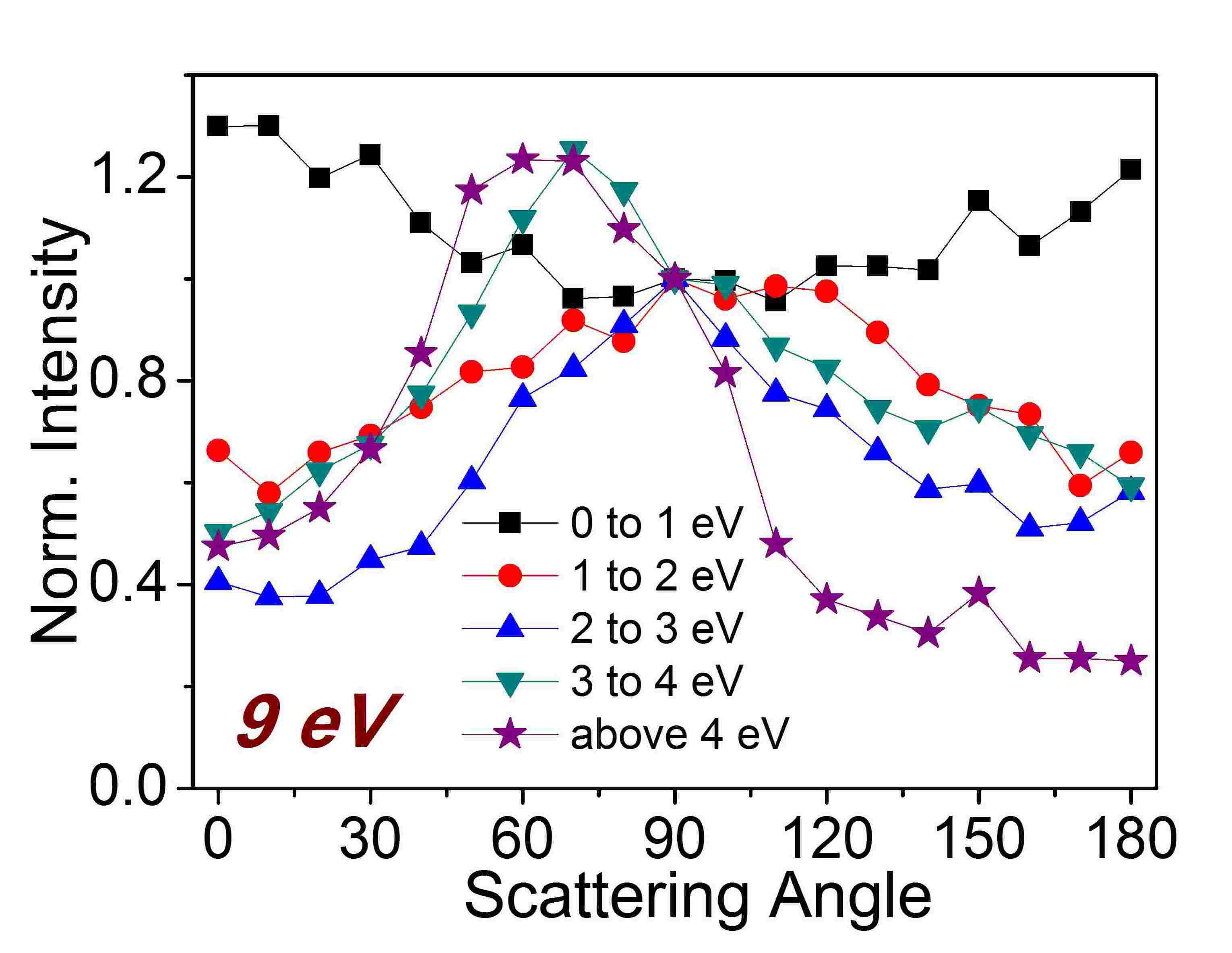}}
\subfloat[10 eV]{\includegraphics[width=0.3\columnwidth]{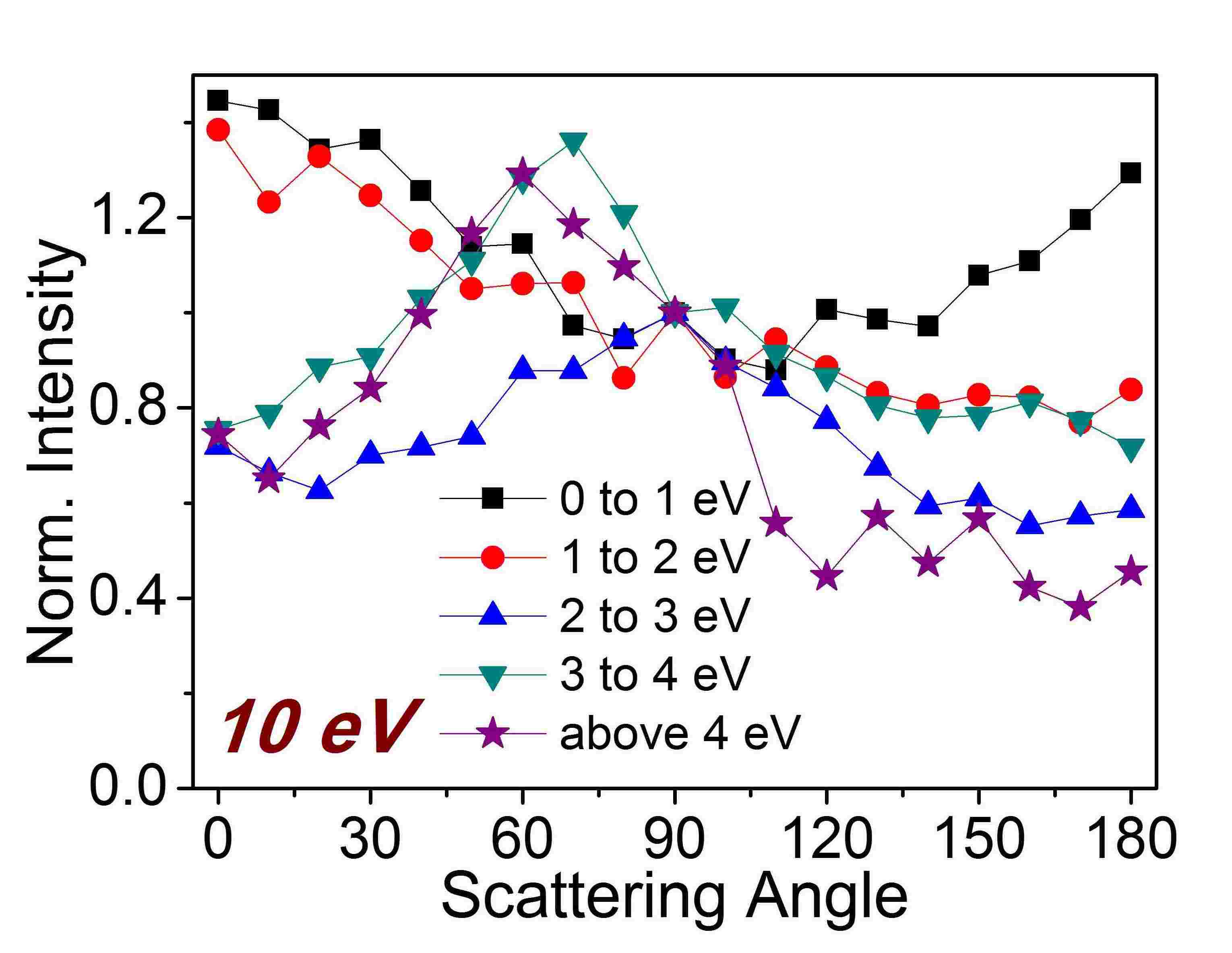}}\\
\subfloat[11 eV]{\includegraphics[width=0.3\columnwidth]{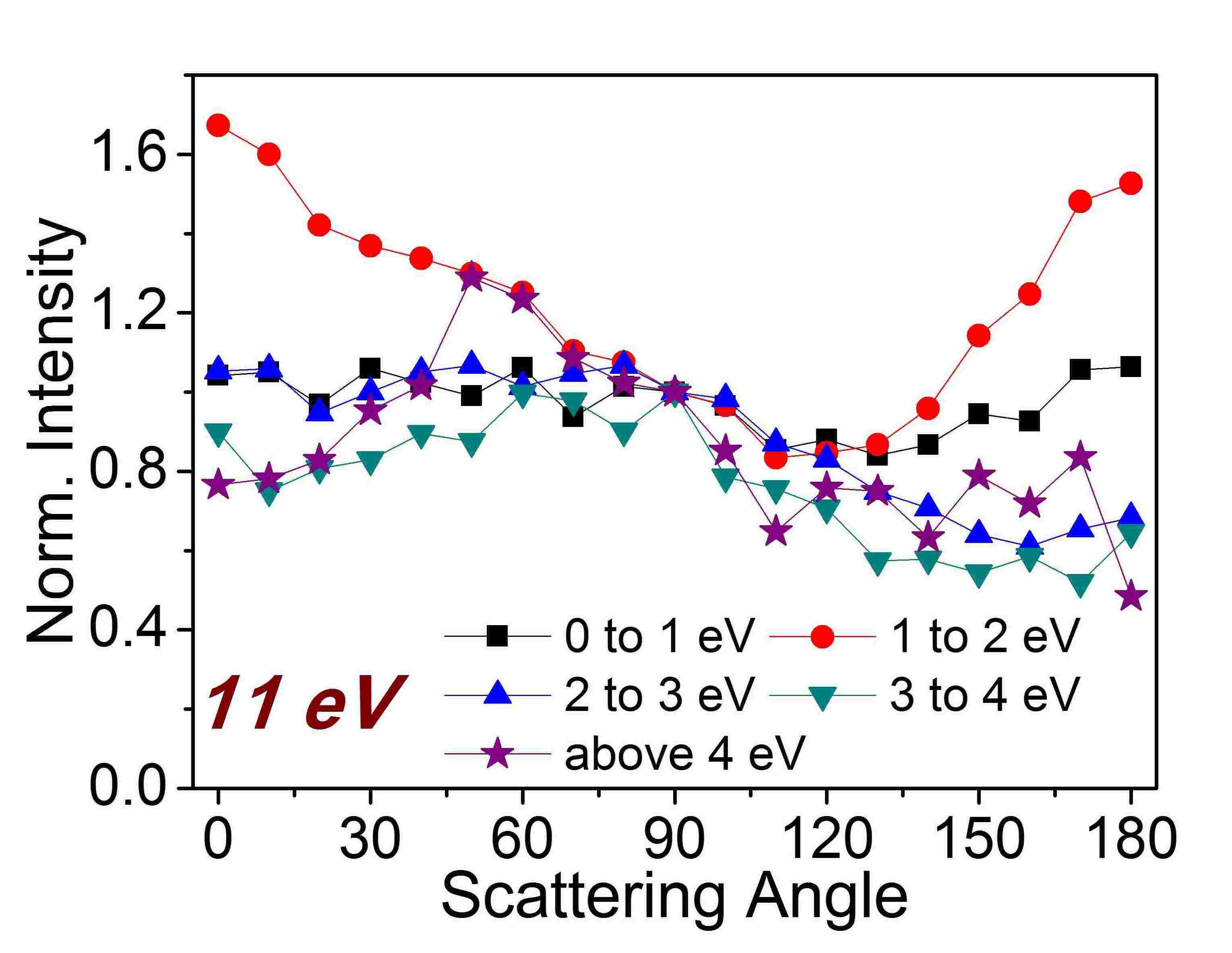}}
\subfloat[12 eV]{\includegraphics[width=0.3\columnwidth]{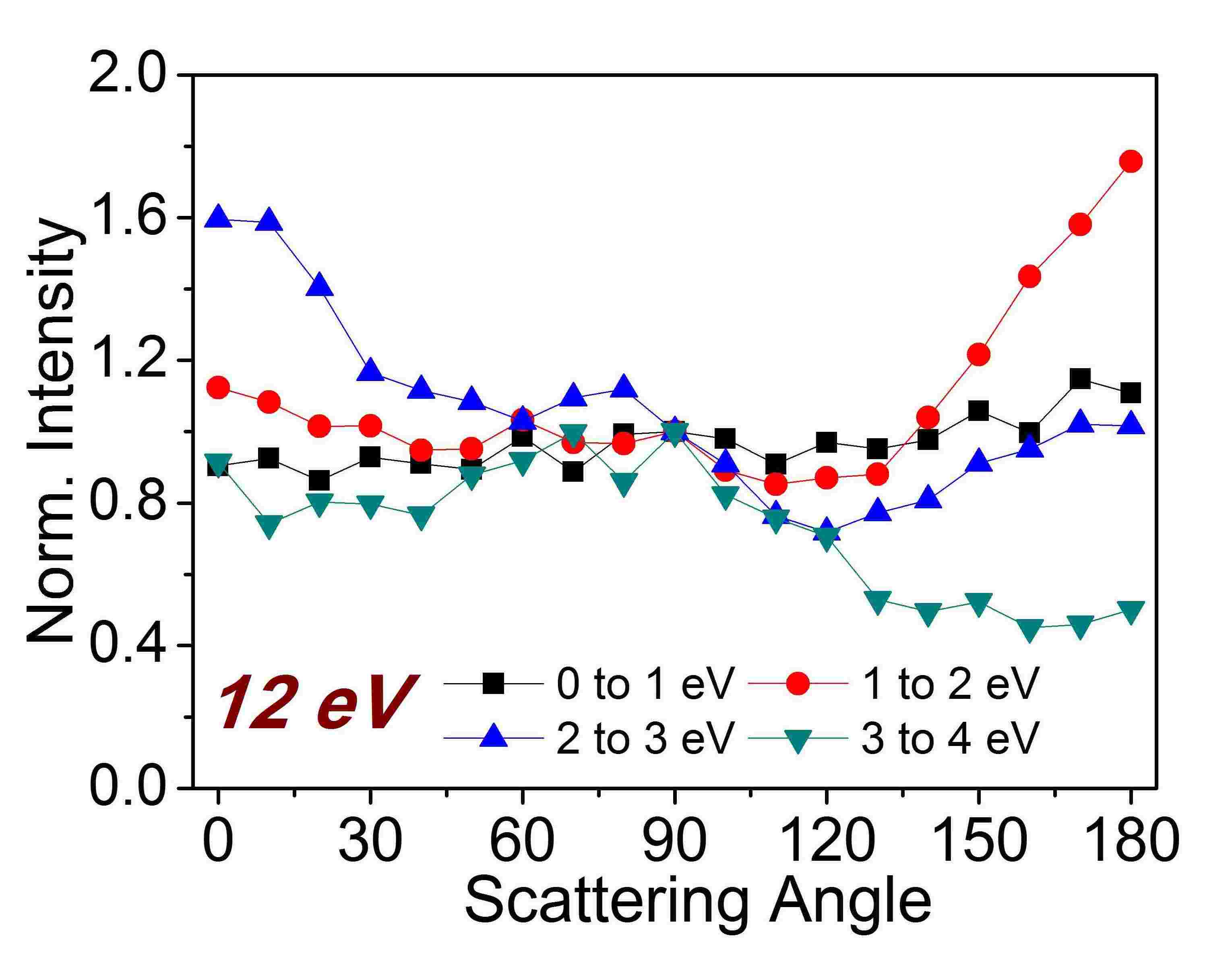}}
\caption{Variation of angular distribution of \ce{H-} with kinetic energy at electron energies (a) 8 eV (b) 9 eV (c) 10 eV (d) 11 eV (e) 12 eV. }
\label{fig5.17}
\end{figure}

We also looked at the variation of the angular distribution of \ce{H-} ions as a function of kinetic energy (Figure \ref{fig5.17}) to infer structural changes of the molecule with internal excitation. For the two-body \ce{H- + CH3} channel dominant at lower electron energies (8, 9 and 10 eV), we see the angular peak of the \ce{H-} ions with KE above 4 eV to be at about $70^{\circ}$ and falling off substantially at forward and backward angles. However, there is a slight asymmetry with the forward angles more intense than the backward angles. For \ce{H-} ions with 3 to4 eV and 2 to 3 eV, the angular distribution widens about the $90^{-}$ direction and the backward angles get more intense. The peaking of the angular distribution close to $90^{\circ}$ shows that most dissociating C-H bonds are perpendicular to the electron beam direction. Thus, the change from a relatively narrow distribution about $70^{\circ}$ to a broad one about $90^{\circ}$ may be due to vibrational excitation/Jahn Teller distortion. Further, the inner group of \ce{H-} ions with lower kinetic energy appears to arise from a different resonance facilitated by the Jahn Teller distortion. Here, the breakup is seen to be a three-body breakup channel with scattering mostly in the forward and backward distribution with a broad dip about the $90^{\circ}$ direction.  

\begin{figure}[!htbp]
\centering
\subfloat[8 eV]{\includegraphics[width=0.3\columnwidth]{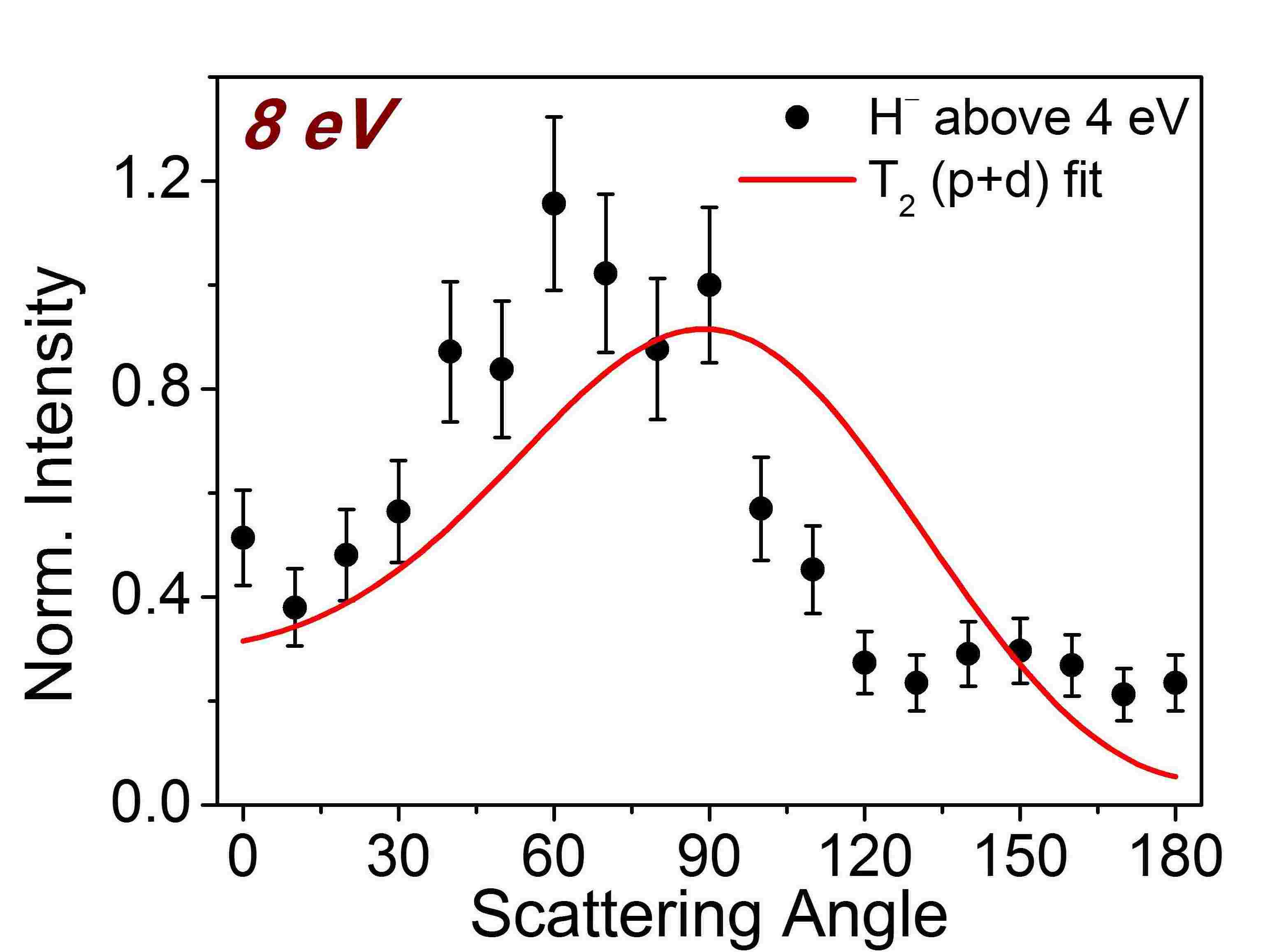}}
\subfloat[9 eV]{\includegraphics[width=0.3\columnwidth]{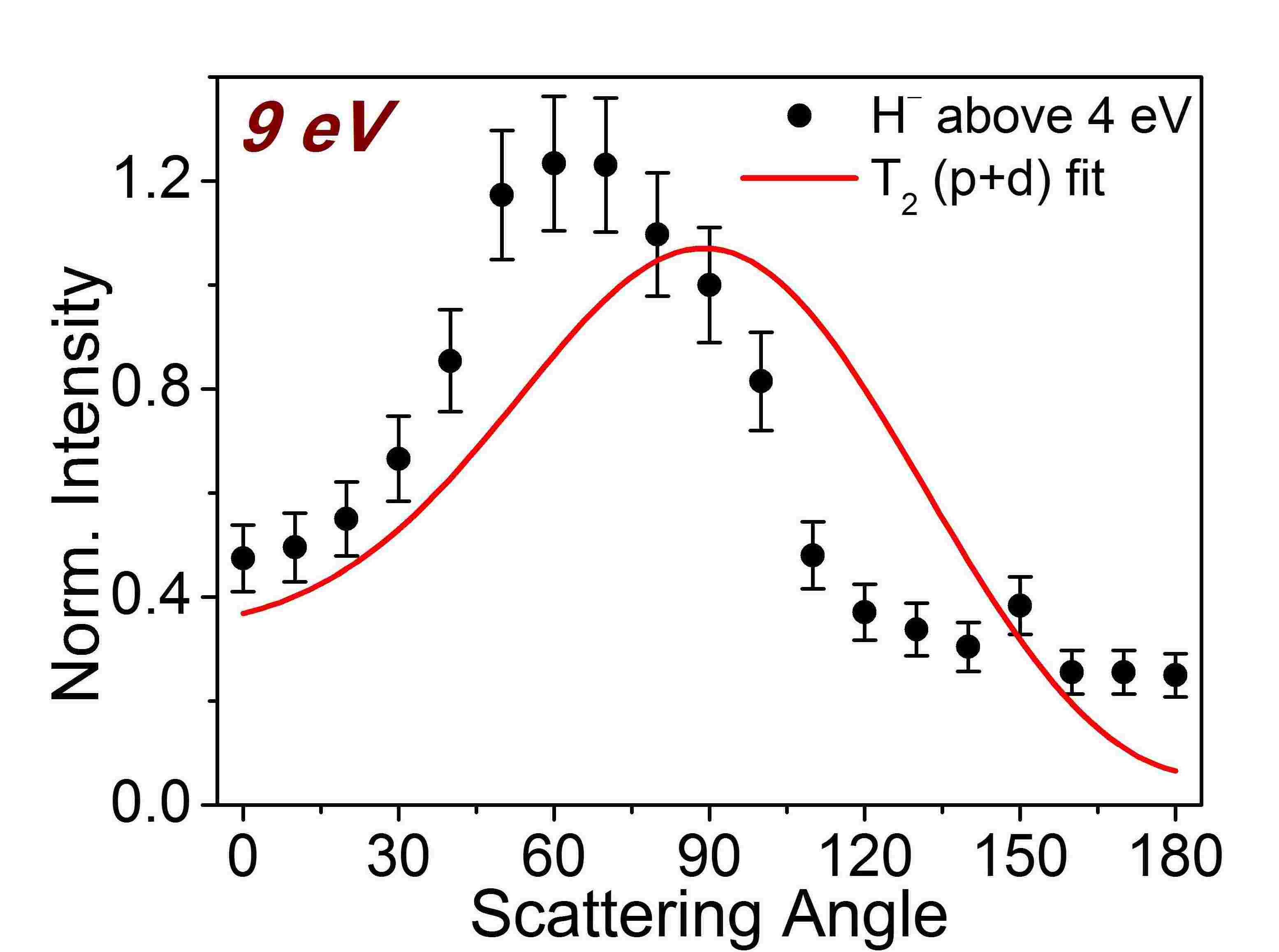}}
\subfloat[10 eV]{\includegraphics[width=0.3\columnwidth]{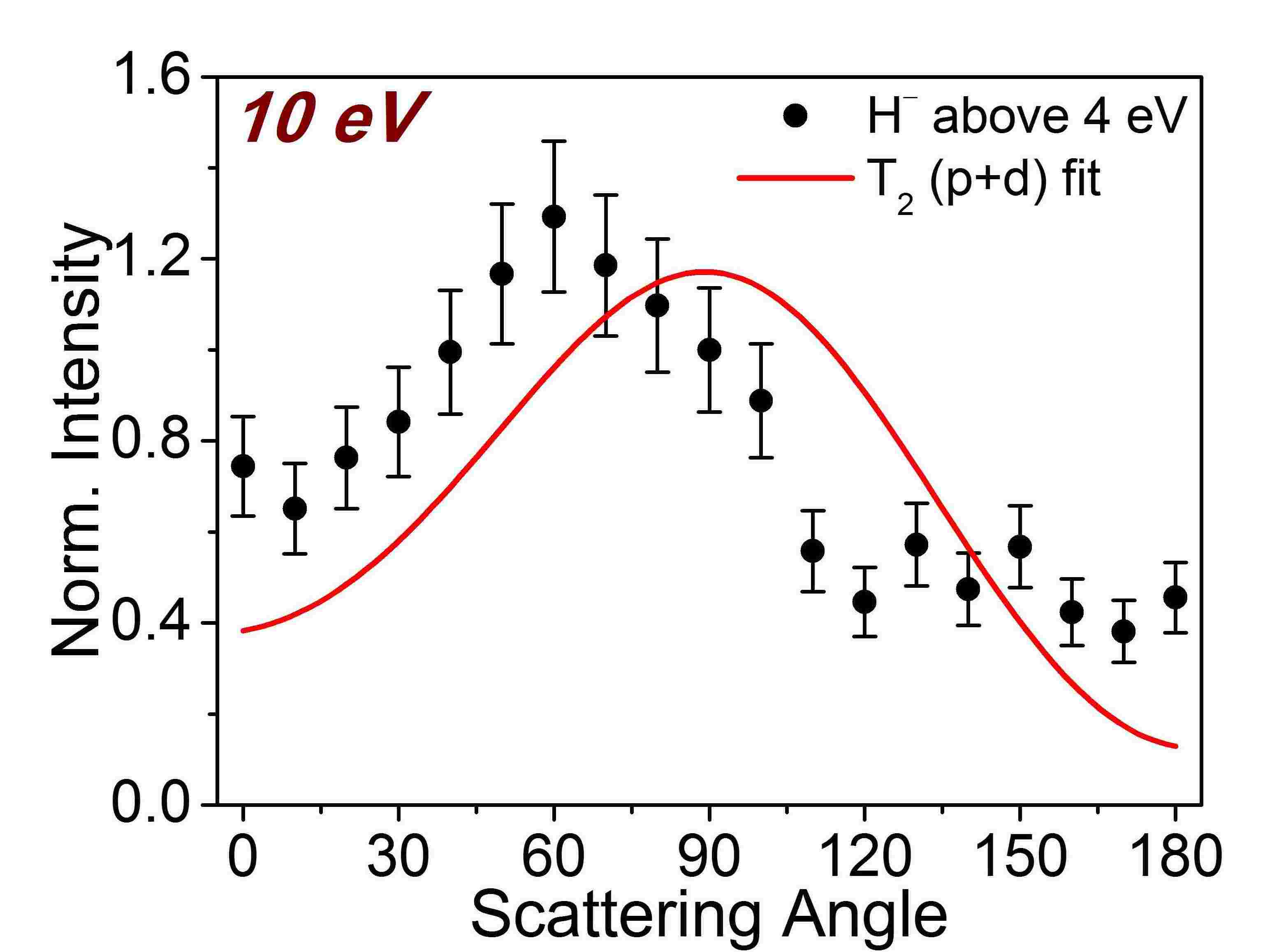}}
\caption{\ce{T2} symmetry fits to angular data of \ce{H-} ions above 4 eV taking lowest allowed $p$ and $d$ partial waves at electron energies 8, 9 and 10 eV. The fit shows reasonable agreement with the data. Since, the electronic symmetry is lowered following the excitation of the triply degenerate \ce{1t2} orbital, \ce{T2} symmetry alone may not account entirely for the observed angular distribution. }
\label{fig5.18}
\end{figure}

We fit the angular distribution of \ce{H-} ions with maximum KE (above 4 eV) arising from the two-body \ce{H- + CH3} breakup channel with \ce{T2} symmetry functions involving lowest allowed $p$ and $d$ partial waves (see Figure \ref{fig5.18}). The fits show a reasonably agreement with a broad peak about $80^{\circ}$ with non zero intensities at the forward and backward. Thus, the highest KE \ce{H-} ions have a dissociation pattern closely determined by the \ce{T2} symmetry under the \ce{T_{d}} geometry. The $p$ and $d$ partial waves are found to in the ratio 1:0.6 with zero phase difference in the fits.  However, a perfect agreement with the \ce{T2} fit is not seen as the highly symmetric \ce{T_{d}} geometry is lowered upon the excitation of the triply degenerate \ce{t2} orbital. 

For the \ce{H-} ions with KE below 2 eV arising from a three body breakup having forward-backward angular distribution, the dissociating CH bond is either parallel or antiparallel to the electron beam. Drawing analogy with the photodissociation and photoelectron experiments, the dissociation could occur via one of the states within \ce{D_{2d}} or \ce{C_{3v}} geometry. Using Wigner-Witmer rules, the \ce{T2} state of methane (under \ce{T_{d}} group) when resolved into species of lower symmetry maps to \ce{B2} or E symmetry within \ce{D_{2d}} group and \ce{A1} or E symmetry with \ce{C_{3v}} group. The inference of the molecular geometry and the bond orientations at the instant of dissociation is complicated by the presence of 4 equivalent CH bonds and the high symmetry of the molecule thereby giving rise to many possibilities. Detailed calculations of the multidimensional potential energy surfaces for the ground and excited states of \ce{CH4} are needed to ascertain that the correlation between the dissociation channels and the parent geometries.

\section{Summary} 

	\begin{enumerate}
		\item First measurement of kinetic energy and angular distribution of fragment anions from DEA to methane.
		\item Broad resonance structure about 10 eV due to electron attachment to \ce{(1t2)^{-1} (3sa1)^{1}} excited state of methane.
		\item Indication of two closely spaced resonances occurring as a consequence of lowering of symmetry  following the lifting of the degeneracy.
		\item Dominant anion fragments are \ce{H-} and \ce{CH2^{-}} produced via \ce{H- + CH3}, \ce{H- + H + CH2}, \ce{CH2^{-} + H2} and \ce{CH2^{-} + 2H} channels. 
		\item Strong role of Jahn Teller effect due to excitation of triply degenerate \ce{1t2} orbital as seen in the angular distributions.  
	\end{enumerate}

\appendix
\newpage
\section{Angular distribution curves for $T_{d}$ point group}
Expression for angular distribution for various symmetries under $T_{d}$ group taking lowest allowed partial waves.

\section*{$A_{1}$ to $A_{1}$ transition}
\begin{eqnarray}
I_{s}(\theta)&=& 1 \\
I_{f}(\theta)&=& \frac{15}{8} \sin^{4}\beta \cos^{2}\beta \cos^{2}\theta (1-\frac{5}{2}\sin^{2}\theta)^{2}
\end{eqnarray}

\section*{$A_{1}$ to $E$ transition}
\begin{eqnarray}
I_{d}(\theta)&=&(\frac{3}{4}\sin^{2}\beta+\frac{1}{2}\sqrt{\frac{3}{2}}(1-\frac{1}{2}\sin^{2}\beta))^{2}\sin^{4}\theta+(\frac{3}{4}-\frac{1}{4}\sqrt{\frac{3}{2}})^{2} \sin^{2}2\beta\sin^{2}2\theta \nonumber \\
&& +2(\frac{1}{4}(3\cos^{2}\beta-1)+\frac{1}{4}\sqrt{\frac{3}{2}}\sin^{2}\beta)^{2}(3cos^{2}\theta-1)^{2} \nonumber \\
&& +\frac{3}{8}(\cos^{2}\beta \sin^{4}\theta + \sin^{2}\beta \sin^{2}2\theta)
\end{eqnarray}

\section*{$A_{1}$ to $T_{1}$ transition}
\begin{eqnarray}
I_{f}(\theta)&=& (-\frac{1}{4}\sqrt{\frac{15}{2}}\cos\beta(1-\frac{1}{4}\sin^{2}\beta)+\frac{1}{4}\sqrt{\frac{15}{2}}\sin\beta(1-\frac{1}{2}\sin^{2}\beta))^{2}\sin^{6}\theta \nonumber \\
&& +(\frac{3}{4}\sqrt{\frac{15}{2}}\sin\beta(1-\frac{1}{2}\sin^{2}\beta)+\frac{1}{2}\sqrt{\frac{15}{2}}\cos\beta(1-\frac{3}{2}\sin^{2}\beta))^{2}\sin^{4}\theta\cos^{2}\theta \nonumber \\ 
&& +(-\frac{3}{4}\sqrt{\frac{15}{2}}\sin^{2}\beta\cos\beta+\frac{1}{4}\sqrt(30)\sin^{2}2\beta)^{2}\sin^{2}\theta(\cos^{2}\theta-\frac{1}{4}\sin^{2}\theta)^{2} \nonumber \\
&& +\frac{30}{8}\sin^{4}\beta\cos^{2}\beta \cos^{2}\theta(1-\frac{5}{2}\sin^{2}\theta)^{2} \nonumber \\
&& + \frac{5}{16}\sqrt{\frac{15}{2}}\sin^{2}\beta-\frac{\sqrt{30}}{16}\sin^{2}2\beta)^{2}\sin^{6}\theta \nonumber \\
&& +(\frac{5}{16}\sqrt{\frac{15}{2}}\sin2\beta-\frac{1}{2}\sqrt{\frac{15}{2}}\cos2\beta)^{2}\sin^{4}\theta\cos^{2}\theta \nonumber \\
&& +(\sqrt{\frac{15}{2}}(\cos^{2}\beta-\frac{1}{4}\sin^{2}\beta)-\sqrt{\frac{15}{2}}\sin\beta(1-\sqrt{\frac{3}{2}}\sin^{2}\beta))^{2}\sin^{2}\theta(\cos^{2}\theta-\frac{1}{4}\sin^{2}\theta)^{2} \nonumber \\
\end{eqnarray}

\section*{$A_{1}$ to $T_{2}$ transition}
\begin{eqnarray}
I_{p}(\theta)&=&2\sin^{2}\theta+2\cos^{2}\beta\cos^{2}\theta+\sin^{2}\beta\sin^{2}\theta \\
I_{d}(\theta)&=& \frac{3}{8}(((1-\frac{\sin^{2}\beta}{2})^{2} + (2\sin\beta + \cos\beta)^{2})\sin^{4}\theta \nonumber \\
&& + (\frac{\sin^{2}2\beta}{4} + (2\cos\beta - \sin \beta)^{2}) \sin^{2}2\theta \nonumber \\
&& + \frac{\sin^{4}\beta}{2} (3\cos^{2}\theta-1)^{2}) \\
I_{f}(\theta)&=&0.39 \sin^{6}\beta \sin^{6}\theta)+ 14.0625 \sin^{4}\beta \cos^{2}\beta \sin^{4}\theta \cos^{2}\theta \nonumber \\
&&+36 \sin^{2}\beta(\cos^{2}\beta-0.25\sin^{2}\beta)^{2}\sin^{2}\theta(\cos^{2}\theta-0.25\sin^{2}\theta)^{2} \nonumber \\ 
&&+2(\cos\beta(1-\frac{5}{2} \sin^{2}\beta) \cos\theta (1-\frac{5}{2} \sin^{2}\theta)+\frac{5}{4\sqrt{2}} \sin^{3}\beta\cos\theta(1-2.5\sin^{2}\theta))^{2}\nonumber \\
&& +\frac{5}{4\sqrt{2}}((\cos^{2}\beta+0.25\sin^{2}\beta)\sin^{3}\theta+\frac{1}{4}\sin^{2}\beta \sin^{3}\theta)^{2} \nonumber \\
&& +\frac{15}{4\sqrt{2}} \sin^{2}\beta \sin\theta(\cos^{2}\theta-0.25\sin^{2}\theta) \nonumber \\
&& +\frac{3}{\sqrt{2}}(\cos^{2}\beta-0.25\sin^{2}\beta)\sin\theta(\cos^{2}\theta-0.25\sin^{2}\theta)^{2} \\
\nonumber \\
I_{p+d}(\theta)&=& a^{2} (2\sin^{2}\theta+2\cos^{2}\beta\cos^{2}\theta+\sin^{2}\beta\sin^{2}\theta) \nonumber \\
&& + \frac{3}{8} b^{2}(((1-\frac{\sin^{2}\beta}{2})^{2} + (2\sin\beta + \cos\beta)^{2})\sin^{4}\theta \nonumber \\
&& + (\frac{\sin^{2}2\beta}{4} + (2\cos\beta - \sin \beta)^{2}) \sin^{2}2\theta  + \frac{\sin^{4}\beta}{2} (3\cos^{2}\theta-1)^{2}) \nonumber \\
&& + \sqrt{1.5} ab (\sin\beta (2\cos\beta-\sin\beta) \sin\theta \sin2\theta  + \cos\beta \sin^{2}\beta \cos\theta (3\cos^{2}\theta-1) \nonumber \\
&&+ \frac{\sin2\beta}{\sqrt{2}} \sin\theta) \cos\delta
\end{eqnarray}

\end{document}